\documentclass{emulateapj}
\usepackage{url}
\usepackage{natbib}
\usepackage{array}

\newcommand{\beq}{\begin{equation}}
\newcommand{\eeq}{\end{equation}}
\newcommand{\bdi}{\begin{displaymath}}
\newcommand{\edi}{\end{displaymath}}
\newcommand{\IRAS}{\textit{IRAS}}
\newcommand{\bolocam}{BOLOCAM}
\newcommand{\irac}{IRAC}

\newcommand{\degree}{$^{\circ}$}

\begin{document}
\shorttitle{The BLAST View of Aquila}
\shortauthors{Rivera-Ingraham, A.~et al.}

\title{The BLAST View of the Star Forming Region in Aquila ($\ell=45$\degree,$\lowercase{b}=0$\degree)}

\author{Alana~Rivera-Ingraham,\altaffilmark{1}
	Peter~A.~R.~Ade,\altaffilmark{2}
        James~J.~Bock,\altaffilmark{3,4}
	Edward~L.~Chapin,\altaffilmark{5}
	Mark~J.~Devlin,\altaffilmark{6}
	Simon~R.~Dicker,\altaffilmark{6}
	Matthew~Griffin,\altaffilmark{2}
	Joshua~O.~Gundersen,\altaffilmark{7}
        Mark~Halpern,\altaffilmark{5}
        Peter~C.~Hargrave,\altaffilmark{2}
	David~H.~Hughes,\altaffilmark{8}
	Jeff~Klein,\altaffilmark{6}
	Gaelen~Marsden,\altaffilmark{5}
        Peter~G.~Martin,\altaffilmark{1,9}
        Philip~Mauskopf,\altaffilmark{2}
	Calvin~B.~Netterfield,\altaffilmark{1,10}
        Luca~Olmi,\altaffilmark{11,12}
	Guillaume~Patanchon,\altaffilmark{13}
	Marie~Rex,\altaffilmark{6}
        Arabindo~Roy,\altaffilmark{1}
        Douglas~Scott,\altaffilmark{5}
	Christopher~Semisch,\altaffilmark{6}
	Matthew~D.~P.~Truch,\altaffilmark{6}
	Carole~Tucker,\altaffilmark{2}
        Gregory~S.~Tucker,\altaffilmark{14}
	Marco~P.~Viero,\altaffilmark{1}
	Donald~V.~Wiebe\altaffilmark{5}}

\altaffiltext{1}{Department of Astronomy \& Astrophysics, University of Toronto, 50 St. George Street, Toronto, ON  M5S~3H4, Canada}
\altaffiltext{2}{Department of Physics \& Astronomy, Cardiff University, 5 The Parade, Cardiff, CF24~3AA, UK}
\altaffiltext{3}{Jet Propulsion Laboratory, Pasadena, CA 91109-8099, USA}
\altaffiltext{4}{Observational Cosmology, MS 59-33, California Institute of Technology, Pasadena, CA 91125, USA}
\altaffiltext{5}{Department of Physics \& Astronomy, University of British Columbia, 6224 Agricultural Road, Vancouver, BC V6T~1Z1,Canada}
\altaffiltext{6}{Department of Physics and Astronomy, University of Pennsylvania, 209 South 33rd Street, Philadelphia, PA 19104, USA}
\altaffiltext{7}{Department of Physics, University of Miami, 1320 Campo Sano Drive, Carol Gables, FL 33146, USA}
\altaffiltext{8}{Instituto Nacional de Astrof{\'i}sica {\'O}ptica y Electr{\'o}nica (INAOE), Aptdo. Postal 51 y 72000 Puebla, Mexico}
\altaffiltext{9}{Canadian Institute for Theoretical Astrophysics, University of Toronto, 60 St. George Street, Toronto, ON M5S~3H8, Canada}
\altaffiltext{10}{Department of Physics, University of Toronto, 60 St. George Street, Toronto, ON M5S~1A7, Canada}
\altaffiltext{11}{University of Puerto Rico, Rio Piedras Campus, Physics Dept., Box 23343, UPR station, San Juan, Puerto Rico}
\altaffiltext{12}{Istituto di Radioastronomia, Largo E. Fermi 5, I-50125, Firenze, Italy}
\altaffiltext{13}{Laboratoire APC, 10, rue Alice Domon et L{\'e}onie Duquet 75205 Paris, France}
\altaffiltext{14}{Department of Physics, Brown University, 182 Hope Street, Providence, RI 02912, USA}

\begin{abstract}
We have carried out the first general submillimeter analysis of the field towards GRSMC 45.46+0.05, a massive star forming region in Aquila. The deconvolved 6 deg$^2$ ($3$\degree$\times2$\degree) maps provided by BLAST in 2005 at 250, 350, and 500 \micron\ were used to perform a preliminary characterization of the clump population previously investigated in the infrared, radio, and molecular maps. Interferometric CORNISH data at 4.8\,GHz have also been used to characterize the Ultracompact \ion{H}{2} regions (UCH{\sc ii}Rs) within the main clumps.
By means of the BLAST maps we have produced an initial census of the submillimeter structures that will be observed by \textit{Herschel}, several of which are known Infrared Dark Clouds (IRDCs). Our spectral energy distributions of the main clumps in the field, located at $\sim7$ kpc, reveal an active population with temperatures of $T\sim$ 35-40\,K and masses of $\sim10^3\,$M$_\odot$ for a dust emissivity index $\beta=1.5$. The clump evolutionary stages range from evolved sources, with extended \ion{H}{2} regions and prominent IR stellar population, to massive young stellar objects, prior to the formation of an UCH{\sc ii}R.
The CORNISH data have revealed the details of the stellar content and structure of the UCH{\sc ii}Rs. In most cases, the ionizing stars corresponding to the brightest radio detections are capable of accounting for the clump bolometric luminosity, in most cases powered by embedded OB stellar clusters.
\end{abstract}

\keywords{ISM: clouds --- balloons --- stars: formation --- submillimeter}


\section{Introduction} \label{sec:intro}
The study of the earliest stages of star formation is intrinsically connected to the ability to probe the dusty regions within which stars are born. The short wavelength radiation from deeply embedded young stellar objects is absorbed by the dust and reaches the observer as re-processed far infrared (FIR) and submillimeter (submm) emission. The low optical depth at these longer wavelengths makes submm studies the best tool to investigate the clumps and inner cores ($\sim1-0.1\,$pc; \citealp{williams2000}) that produce new stars, especially the `coldest' structures prior to the onset of stellar activity.

The 2-m Balloon-borne Large-Aperture Submillimeter Telescope (BLAST; \citealp{pascale2008}), with two consecutive and successful science missions in 2005 and 2006, mapped the sky simultaneously at the crucial wavelength bands of 250, 350, and 500\,\micron. The Galactic (e.g., \citealp{netterfield2009}) and extragalactic (e.g., \citealp{devlin2009}) data provided by the BLAST mission demonstrate the exciting science to be continued with the \textit{Herschel Space Observatory}\footnote{http://www.esa.int/SPECIALS/Herschel/index.html}(hereafter \textit{Herschel}; e.g., \citealp{elia2010}) and its SPIRE instrument (\citealp{griffin2009}; $18$\arcsec\ at 250\,\micron).

In this study we present a preliminary analysis of the 6 deg$^2$ field of a massive star forming region in Aquila ($\ell=45$\degree, $b=0$\degree), observed by BLAST during its 2005 flight. This work therefore complements the analysis of the other BLAST05 Galactic plane star forming regions, including Vulpecula (\citealp{chapin2008}) and Cygnus X (Roy et al. in preparation).

Aquila is well known for its prominent massive star formation (e.g., \citealp{rathborne2004}), and is dominated by two main molecular complexes. These are host to numerous Ultracompact \ion{H}{2} regions (UCH{\sc ii}Rs; e.g., \citealp{wood1989}; [WC89]), maser emission, and outflows from young and highly embedded OB stellar clusters. The stellar activity of the Galactic Ring Survey Molecular Cloud (GRSMC) 45.60+0.30 has been extensively studied in the infrared and radio (e.g., \citealp{kraemer2003}; \citealp{vig2006}). However, although submm data have also been used to complement the detailed study of this cloud (e.g., \citealp{vig2006}), the overall population and properties of this entire region in Aquila are yet to be fully characterized at submm wavelengths. 

Previous BLAST studies of regions like Vulpecula \citep{chapin2008} and Vela \citep{netterfield2009} have provided robust samples of submm sources in a range of evolutionary stages.  By fitting modified blackbody spectral energy distributions (SEDs) to the BLAST measurements, and with the aid of ancillary data at infrared (IR) wavelengths for the Vulpecula sources (including \IRAS, \textit{Spitzer} MIPS, and the \textit{Midcourse Space Experiment} (\textit{MSX})), these studies constrained dust temperatures, masses and bolometric luminosities for their respective samples. Their results aimed to characterize and investigate crucial properties of the submm population, including their star formation stage, the (core) mass function, and core lifetimes (Vela). 

In this paper we provide the first general characterization of the submm clump population of Aquila, which will serve as the initial census and preliminary analysis of those structures that will be observed in more detail by \textit{Herschel} during its surveys of the massive star forming regions in the Galactic Plane (Hi-GAL: The \textit{Herschel} infrared Galactic Plane Survey; \citealp{higal2010}). Like in previous BLAST studies, we provide estimates of physical parameters such as mass, dust temperature, and luminosity, although in this work this analysis is constrained to the most prominent clumps in the field, lying at the common distance of $\sim7$\,kpc (\S~\ref{distest}). We have complemented our submm analysis with a detailed study of the UCH{\sc ii}Rs and OB clusters within these main clumps with the aid of 4.8\,GHz radio interferometry data provided by the Co-Ordinated Radio `N' Infrared Survey for High-mass star formation (CORNISH; \citealp{purcell2008}). In \S~\ref{sec:obs} we describe the submm and radio data used in this work, as well as the techniques required to produce our final results. We then analyze the major clouds and clumps in the region (\S~\ref{sec:sources}), whose properties we discuss in \S~\ref{sec:discussion}. In \S~\ref{sec:end} we give a brief summary of our results and the prospects for future studies of this massive star forming region.

\section{Observations \& Data Analysis} \label{sec:obs}
\subsection{Submillimeter Data Products and Source Detection}\label{dproducts}
BLAST observed Aquila during its first science flight in June 2005, mapping a total of $6$ deg$^2$ ($3$\degree$\times2$\degree) at 250, 350, and 500\,\micron\ towards GRSMC 45.60+0.30.
For the following analysis we note that the three BLAST maps were made such that the average of these maps is zero \citep{patanchon2008}. During the reduction and mapmaking process, the SANEPIC (Signal and Noise Estimation Procedure) algorithm removed low-frequency noise (mainly sky signal). Large scale signals in the map were recovered, but the DC level was set to zero by application of a high-pass filter to the time-ordered data. The reduction process, as well as the subsequent procedures to correct the unexpected optical degradation that occurred during the flight \citep{truch2008}, have been described in detail in \citet{chapin2008} and \citet{roy2010}. The effective resolution is $\sim$40, 50, and 60\arcsec\ at 250, 350 and 500\,\micron, respectively. The 250 and 350\,\micron\ maps were later convolved in order to produce BLAST maps with a common resolution of $60$\arcsec\ for photometry analysis.

\begin{figure*}[ht]
\centering
\includegraphics[scale=0.64,angle=270]{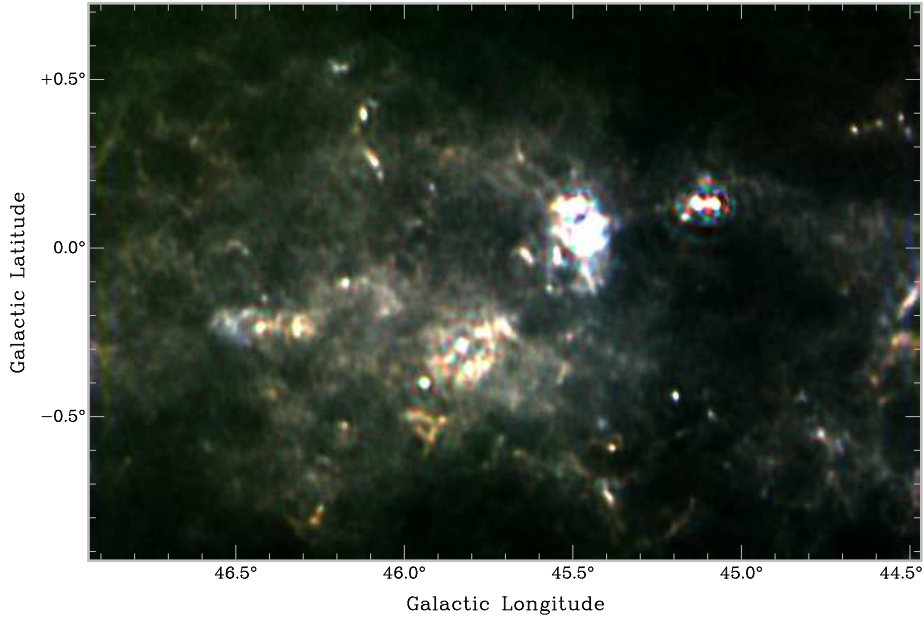}
\caption{BLAST submm map in Aquila combining three wavebands at 250, 350 and 500\,\micron, with color coding blue, green and red, respectively.}
\label{fig:aquila}
\end{figure*}

As can be seen in Figure \ref{fig:aquila}, the final products still suffered from artifacts (ripples) surrounding the brightest objects and the map edges arising from the deconvolution process. These introduce false detections when running the Interactive Data Language ({\sc idl}\footnote{{\sc IDL} is a product of ITT Visual Information Solutions, http://ittvis.com/}) routines designed to detect the main submm peaks in the field, and which we therefore have to eliminate manually after visual inspection.
For source identification we used the IDL-implemented DAOPHOT `{\sc find}' routine and a S/N detection limit of $3$ at 250\,\micron, the band with higher signal to noise ratio (S/N) for a typical star-forming clump. We kept a total of 66 sources between $\ell=44$\degree$.6$ and $46$\degree$.8$ (to avoid map edges) which also had a match at 350\,\micron\ within 20\arcsec\ with a S/N$>3$. We also kept six additional sources (identified with (*) in Table \ref{table:srcs}) with a lower S/N at 250 and 350\,\micron\ (combined S/N between $\sim2-3.5$, visually prominent, and some with counterparts in the IRAS Point Source Catalog; PSC). Their lower S/N can be explained by their location in the local neighborhood of (clustered) bright sources, regions with difficult background estimation and significantly above average noise. This source catalog (Table \ref{table:srcs}) is robust to $\sim50\,$Jy at 350\,\micron\ ($\sim140$\,Jy at 250\,\micron). For comparison, the cirrus noise ($3$\,$\sigma_{\mathrm{cirrus}}$) at 250, 350 and 500\,\micron\ is estimated to be $\sim20$, $10$, and $5$\,Jy for each map, respectively \citep{roy2010}. In the present work we did not require completeness to faint flux densities, however, choosing to focus our analysis only on the most prominent submm peaks of our sample. 

\begin{figure*}[ht]
\centering
\includegraphics[scale=0.48,angle=270]{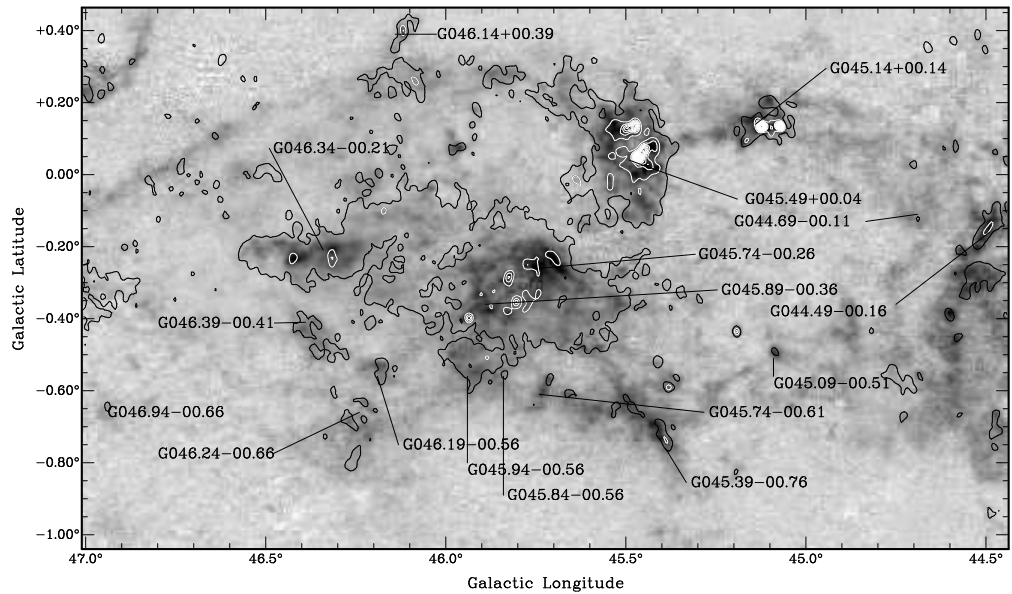}
\caption{Greyscale: $^{13}$CO molecular emission at $\sim60\pm10$\,km s$^{-1}$ with labels marking position and names (GRSMC) of clouds in this range from \citet{rathborne2009}. Contours are BLAST 500\,\micron\ surface densities. White: contours from 10 to 90\% of map peak value of 2153\,MJy\,sr$^{-1}$ in 5\% steps. Black: 5\% contour level (see text).}
\label{fig:molecular}
\end{figure*}

\begin{figure}[ht]
\centering
\includegraphics[scale=0.5]{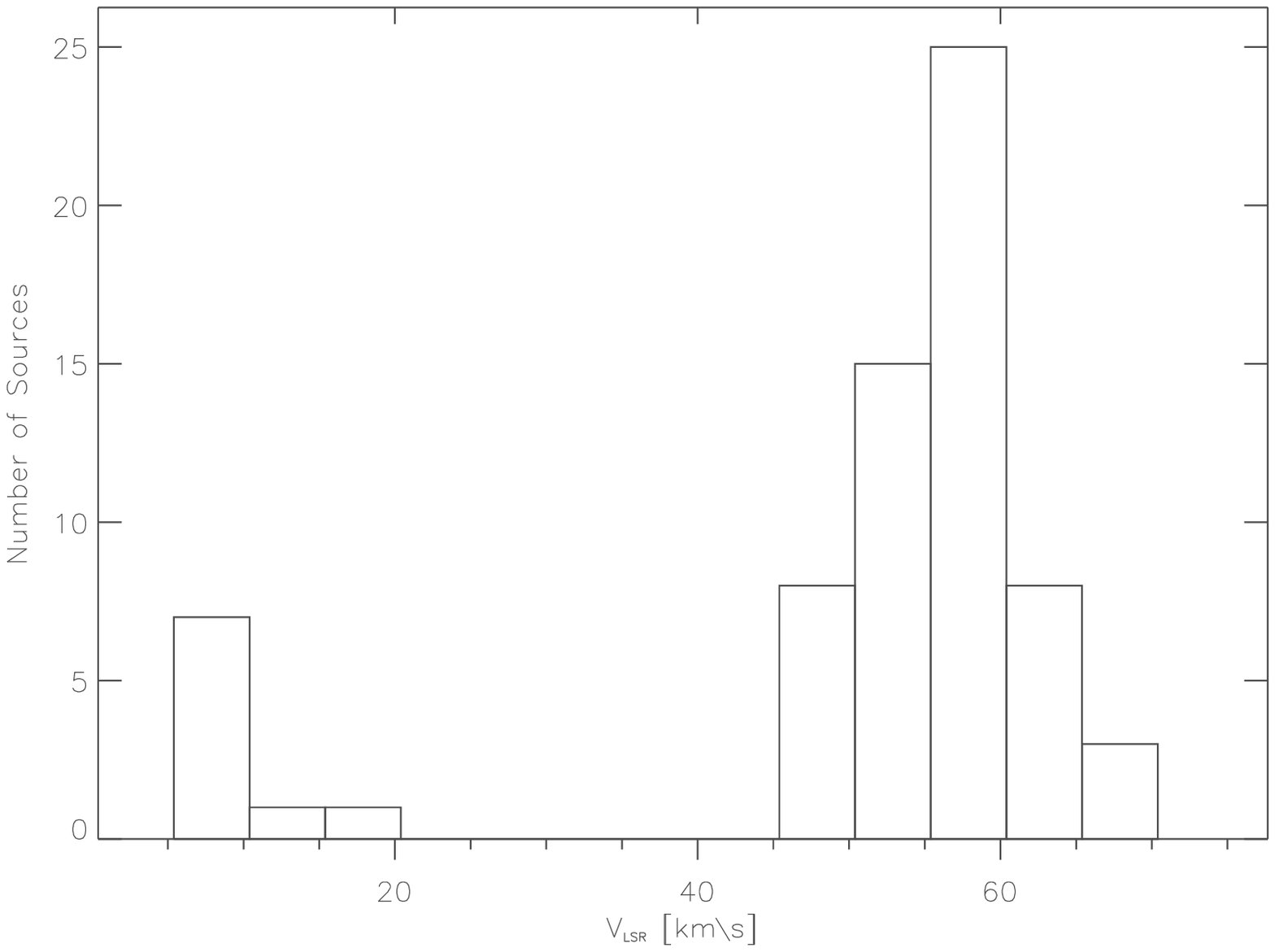}
\caption{Histogram of $^{13}$CO radial velocities of 68 BLAST detections with molecular clump counterpart (BLAST peak lying within spatial FWHM of clump) in catalogs of \citet{roman2009} and \citet{rathborne2009}. The main stellar activity has velocities within the broad (high velocity) peak.}
\label{fig:vels}
\end{figure}

\subsection{Distance Estimation}\label{distest}
In order to investigate the velocities and distances towards the main clumps in the field we used the $^{13}$CO molecular spectra from the BU-FCRAO Galactic Ring Survey (GRS; \citealp{grs2006}). The datasets cover the region between $18$\degree$<\ell<55$\degree$.7$ with a velocity resolution of 0.2 km\,s$^{-1}$. The spatial resolution is 46\arcsec\ with an angular sampling of 22\arcsec. Due to the distribution of the clumps throughout the field, this spectral analysis is essential in order to estimate and justify the use of a suitable common distance to the most active regions in Aquila. A comparison of BLAST with the molecular emission is shown in Figure \ref{fig:molecular}. To aid qualitative intercomparison, BLAST contours are presented in this and a number of later figures, from images in which an approximate DC level has been restored (300, 160, and 50\,MJy\,sr$^{-1}$ at 250, 350, 500\,\micron, respectively).

For the spectral analysis we first extracted and fitted the $^{13}$CO spectra at the position of each submm peak. Molecular counterparts from the catalogs of \citet{roman2009} and \citet{rathborne2009} were assigned to each BLAST peak if the peak was located within the spatial FWHM of the molecular clump. The velocity we obtained from the spectral fitting was then compared with the clump velocities in the catalog. This allowed us to distinguish the appropriate molecular counterpart to the BLAST clump in those cases where more than one molecular counterpart (but with different Local Standard of Rest (LSR) radial velocity) was found for a particular BLAST clump (e.g., Table \ref{table:srcs}).
Our analysis and line fitting of the spectra is in good agreement with the significant dispersion of velocities (e.g., Figure \ref{fig:vels}) existing in the recent cloud/clump catalogs of \citet{rathborne2009} and \citet{roman2009}. The main activity of the field is located in the main broad high-velocity peak in Figure \ref{fig:vels}. The smaller peak at lower velocities could arise from clumps located either much closer or much farther away ($\sim1$ or $\sim11$\,kpc; \citealp{roman2009}) than the complexes forming the high-velocity peak (which cluster around the tangent point of $\sim6$\,kpc; e.g., \citealp{simon2001}).

Using the \citet{clemens1985} rotation curve of the Milky Way, \citet{roman2009} estimated distances to GRS molecular clouds (e.g., \citealp{rathborne2009}) with the aid of \ion{H}{1} self-absorption (HISA) and continuum absorption, both in data from the Very Large Array Galactic Plane Survey (VGPS; \citealp{stil2006}). There are four clouds in their sample within the defined BLAST map limits (\S~\ref{dproducts}) with an assigned distance between $\sim7\pm0.5$\,kpc (LSR velocity within the second peak in Figure \ref{fig:vels}: v$\sim60$ km\,s$^{-1}$): G045.49+00.04, G045.74$-$00.26, G045.89$-$00.36, and G045.39$-$00.76. However, one of the major clouds in the BLAST field, G045.14+00.14 (v$_{\mathrm{LSR}}\approx60$ km\,s$^{-1}$), has instead been assigned its near distance of $\sim4.5$\,kpc.

Taking into account a maximum standard error of $10$\% for the distances \citep{roman2009}, we estimate that a common distance of $\sim7$\,kpc would be a suitable approximation for the analysis of all the clouds with velocities of $\sim60$\,km s$^{-1}$ and located in this distance range. Furthermore, this estimate is in good agreement with the distances derived by \citet{pandian2009}, who also resolved the kinematic distance ambiguities using \ion{H}{1} self-absorption with data from the VGPS. These authors calculated, by means of methanol masers and a flat rotation curve, a range of distances for the clumps within G045.49+00.04 between $6.9$ and $7.4$\,kpc, which again agrees well with the estimate from \citet{roman2009}. However, they assigned the far distance of $7.4$\,kpc to the clump in G045.14+00.14, which disagrees with \citet{roman2009}. This coincides with previous studies (e.g., \citealp{simon2001}) that placed this cloud at the same kinematic distance as its neighbor G045.49+00.04.
In this paper we adopt the far distance estimate of \citet{pandian2009} for G045.14+00.14, and we restrict our main analysis to four clouds (G045.49+00.04, G045.74$-$00.26, G045.89$-$00.36, and G045.14+00.14), which comprise the most active complexes of the field. Our main BLAST detections corresponding to G045.39$-$00.76 are associated with a bubble \citep{churchwell2006}, separate from the main clouds of the field, and we therefore omit these sources from the main analysis. A common distance of $\sim7$\,kpc ($1$\degree$\sim0.1$\,kpc scale) is assumed in the present study (in contrast with those used by previous authors of $\sim6$ and $\sim8$\,kpc; e.g., \citealp{simon2001}; \citealp{kraemer2003}; \citealp{paron2009}).

\subsection{Ancillary Data}
To create our SEDs we used the BLAST fluxes estimated using our own Gaussian fitting routine (\S~\ref{photo}). Whenever possible, we also included data from the \IRAS\ Galaxy Atlas (IGA; \citealp{cao1997}: HIRES; \citealp{hires1990}) at 60 and 100\,\micron\ and \bolocam\ (Galactic Plane Survey) data at 1.1\,mm \citep{bolocam2010}. \IRAS\ flux densities were estimated individually using our Gaussian fitting routine, while for the convolved (60\arcsec) \bolocam\ images we used aperture photometry (single sources: an aperture centered on the BLAST source with 1\arcmin.8 radius; an estimate for the background was obtained from an annulus formed by the source aperture and another aperture with radius 1.3 times larger than that of the source). The measured \bolocam\ fluxes were found to be within the error range of those provided by the GATOR service\footnote{http://irsa.ipac.caltech.edu/applications/Gator/}, and the latter values were therefore used to constrain the SEDs at longer wavelengths if a counterpart to the BLAST clump was found within 30\arcsec\ (we found 39 sources in Table \ref{table:srcs} with \bolocam\ counterparts, including all those with measured SEDs, with the exception of A22). We note that \IRAS\ was found to be saturated for the brightest clumps, for which we adopted the upper limits estimated by \citet{kraemer2003}. Most of our measured \IRAS\ fluxes were also found to be higher than those from the \IRAS\ PSC. This effect, combined with the lower resolution of \IRAS\ and the photometry of sometimes blended sources, imply that these values should also be generally considered upper limits in our fits.

To investigate the stellar content and the morphology of the emission at shorter wavelengths we used the InfraRed Array Camera (\irac) on \textit{Spitzer}. Of the four available \irac\ bands ($3.6$, $4.5$, $5.8$, and $8$\,\micron), we focused on the longest wavelength in our morphology analysis. For our study of the stellar population we used data from the Galactic Legacy Infrared Midplane Survey Extraordinaire (GLIMPSE; \citealp{glimpse1}) through the online {\sc gator} query system. The GLIMPSE I online datasets, covering about 220 deg$^2$ of the Galactic Plane ($\ell=10$\degree$-65$\degree) in all four \irac\ bands, include a highly reliable (GR) catalog and a more complete, but less reliable one (GC). Both were used in the present work. 

\begin{figure}[ht]
\includegraphics[scale=0.45]{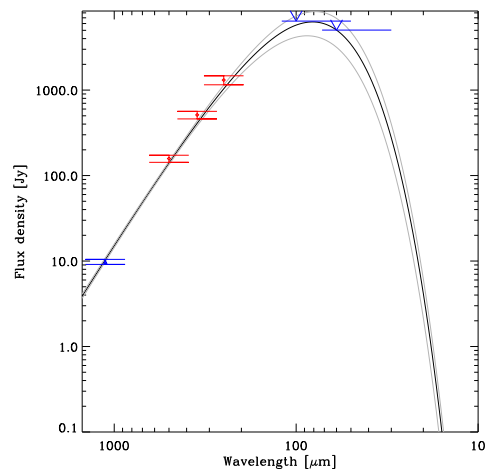}
\caption{Spectral energy distribution of A16 (Table \ref{table:srcs}). Symbols are BLAST and ancillary data (\IRAS; \bolocam). Upper and lower curves are the 1-sigma envelope of the dust emission model \citep{chapin2008}.}
\label{fig:sed1}
\end{figure}

\subsection{Photometry: Spectral Energy Distributions and Parameter Estimation}\label{photo}
Similar to other Galactic BLAST studies (e.g., \citealp{chapin2008}), our SED analysis was carried out by estimating flux densities with our own {\sc idl} spatial Gaussian-fitting routine, and subsequently fitting them with a modified blackbody of the form:

\begin{figure}[ht]
\includegraphics[scale=0.45]{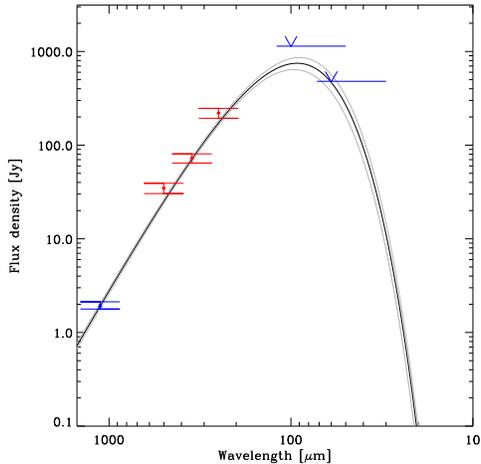}
\caption{Same as Fig.\ref{fig:sed1}, but for A50 (Table \ref{table:srcs}).}
\label{fig:sed2}
\end{figure} 

\begin{equation}
	S_\nu=A\left(\frac{\nu}{\nu_0}\right)^\beta B_\nu(T), 
\end{equation}
where $\beta$ is the dust emissivity index, ($\nu/\nu_0$)$^\beta$ is the emissivity factor normalized at a fixed frequency $\nu_0$, and $B_\nu$($T$) is the Planck function. The amplitude $A$ can be expressed as:

\begin{equation}
	A=\frac{M_{\mathrm{c}}\kappa}{RD^2}, 
\end{equation}
 where $D$ is the distance to the source, $M_{\mathrm{c}}$ is the total clump mass, $\kappa$ is the dust mass absorption coefficient, evaluated at $\nu_0=c/250$\,\micron, and $R$ is the gas-to-dust ratio. A 2-D Gaussian is fitted to the three bands simultaneously with the amplitudes, centroid, orientation angle, and major and minor Full Width at Half Maximum (FWHM) as free parameters. The fit is carried out using a non-linear least-squares minimization {\sc idl} routine (based on {\sc mpfit}: \citealp{mpfit}), which also allowed for simultaneous multiple Gaussian fitting in crowded regions of the field. The background was estimated by fitting a fourth-order polynomial to the regions around the clump.

Following \citet{chapin2008}, estimates and uncertainties for the dust temperatures, masses, and luminosities were obtained through a Monte Carlo analysis performed on our SED models, keeping the dust emissivities and $R$ as fixed parameters ($\beta=1.5$, $R=100$, $\kappa=10$\,cm$^{2}$g$^{-1}$). Examples of SEDs are shown in Figure \ref{fig:sed1} and Figure \ref{fig:sed2}. A flux density consistency check between all convolved and deconvolved maps shows flux conservation to better than 5\%.

The bolometric luminosities derived from the fits for each evolved clump were subsequently used to estimate equivalent single zero-age-main-sequence star (ZAMS) spectral types using the tables of \citet{panagia1973}. This luminosity value was used as a general measure of the energy output of each submm detection. We note that the BLAST sources at a distance of $\sim7$\,kpc are pc-sized, and therefore `clump-like' objects (e.g., \citealp{bergin2007}). We expect them to be powered by several embedded stars, with the luminosity dominated by that of the most massive source(s) within the clump. In addition, we also include the equivalent luminosity class V star spectral types from the more recent stellar calibration provided by \citet{martins2005}, which takes into account the effects of line-blanketing and winds in the non-LTE atmosphere models of O stars. For this case we used the calibration based on the observational $T_{\mathrm{eff}}$ scale; these authors note that the theoretical and observational scales are in good agreement for early type dwarfs and supergiants.

\subsection{4.8\,GHz Radio Interferometry Data and Analysis}
For our radio analysis we used the CORNISH interferometry datasets, a 4.8\,GHz VLA survey covering the northern GLIMPSE region with a resolution of $\sim1$\arcsec. Our analysis of these data was carried out using the same procedure and Gaussian fitting codes used for the submm. The 4.8\,GHz fluxes from our fitting were then used to estimate the total number of ionizing photons per unit time for each radio source (e.g., \citealp{panagia1973}; \citealp{martins2005}):

\begin{equation}
	Q_0=\int^{\infty}_{\nu_0}\frac{L_\nu}{h\nu}d\nu ,
\end{equation}
where $\nu_0$ is the frequency at the Lyman edge.

These values were also used in conjunction with the tables from \citet{panagia1973} and \citet{martins2005} to estimate the equivalent single-star radio spectral types for each detection. The positions and parameters derived using these data have been included in Tables \ref{table:rsrcs}, \ref{table:radio}, and \ref{table:radio1}. The target S/N of our sample is $>3$. As with the BLAST sources, we have also kept a few radio detections with lower S/N. These were identified and required by our Gaussian fitting routine, and are almost exclusively secondary peaks neighboring the brightest sources (clusters: A16-C24, A28-C15), where the S/N is greatly deteriorated ($>1.5$). This effect has already been discussed in \citet{purcell2008}, and these sources will therefore require further examination as soon as additional/better data becomes available. Extra detections with low S/N have been identified (*) in Table \ref{table:rsrcs}. Nomenclature of radio sources is formed by the name of the BLAST clump, followed by that of the CORNISH detection (A\#$-$C\#).

Considering the uncertainties in our measurements and the above procedure for spectral type estimation, assuming optically thin emission, we estimate an uncertainty for these measurements of $\sim0.5$ in spectral type (for a given calibration scale of the stellar parameters). This uncertainty is expected to be larger than any spectral type error that could arise when comparing our estimates (for a distance of $\sim7$\,kpc) and those obtained by previous authors with a distance of $\sim6$ kpc (Tables \ref{table:radio} and \ref{table:radio1}).

\setlength{\extrarowheight}{1.5pt}
\begin{longtable*}{llccllllllc}
\caption[Submm clumps in the BLAST Aquila map.]{Submm clumps in the BLAST Aquila map.} \label{table:srcs} \\
\hline \hline \\
  \multicolumn{1}{c}{Source} &
  \multicolumn{1}{c}{BLAST Name\tablenotemark{a}} &
  \multicolumn{1}{c}{$\ell$} &
  \multicolumn{1}{c}{$b$} &
  \multicolumn{1}{c}{\IRAS\ Name\tablenotemark{b}} &
  \multicolumn{1}{c}{IRDC\tablenotemark{c}} &
  \multicolumn{1}{c}{Cloud\tablenotemark{d}} &
  \multicolumn{1}{c}{Clump\tablenotemark{e}} &
  \multicolumn{1}{c}{$D$ (kpc)\tablenotemark{f}} &
  \multicolumn{1}{c}{Assoc.\tablenotemark{g}} \\ \hline 
  \\
\endfirsthead

\multicolumn{10}{c}{{\tablename} \thetable{} -- Continued} \\
\hline \hline \\ 
  \multicolumn{1}{c}{Source} &
  \multicolumn{1}{c}{BLAST Name} &
  \multicolumn{1}{c}{$\ell$} &
  \multicolumn{1}{c}{$b$} &
  \multicolumn{1}{c}{\IRAS\ Name} &
  \multicolumn{1}{c}{IRDC} &
  \multicolumn{1}{c}{Cloud} &
  \multicolumn{1}{c}{Clump} &
  \multicolumn{1}{c}{$D$ (kpc)} &
  \multicolumn{1}{c}{Assoc.} \\ \hline 
  \\
\endhead

  \hline \\ \multicolumn{10}{l}{{Continued on Next Page\ldots}} \\
\endfoot

\endlastfoot
A0&J191131+102635&$44.502$&$0.347$&19091+1021&&&&&\\
A1&J191124+102845&$44.522$&$0.387$&19090+1023&&&&&\\
A2&J191432+100817&$44.577$&$-0.454$&&P3726&G044.34-00.21&c12&4.12&Y\\
A3&J191135+103138&$44.585$&$0.370$&&&&&&\\
A4&J191148+103504&$44.660$&$0.350$&19094+1029&&&&&\\
A5&J191334+102308&$44.686$&$-0.129$&19112+1018&&G044.49-00.16&c8&7.55&\\
&&&&&&G044.69-00.11&c1&3.7&Y\\
A6&J191514+101158&$44.711$&$-0.577$&&&G045.09-00.51&c4&6.0&\\
&&&&&&G044.74-00.56&c1,c7&$-$&Y\\
A7&J191537+101015&$44.730$&$-0.675$&&&G044.74-00.56&c2&$-$&Y\\
A8&J191514+101518&$44.760$&$-0.551$&&&G044.74-00.56&c1,c8&$-$&Y\\
A9&J191457+102119&$44.816$&$-0.442$&19125+1015&&G045.09-00.51&c6&6.0&\\
A10&J191611+101914&$44.927$&$-0.728$&&&&&&\\
A11&J191606+102731&$45.039$&$-0.644$&&&&&&\\
A12&J191321+105049&$45.070$&$0.132$&19110+1045&&G045.14+00.14&c1&4.53&Y\\
A13&J191538+103419&$45.087$&$-0.492$&19132+1029&&G045.09-00.51&c1&6.0&Y\\
A14&J191528+103636&$45.100$&$-0.435$&&&&&&\\
A15&J191421+104622&$45.118$&$-0.119$&19119+1041&&&&&\\
A16&J191327+105338&$45.123$&$0.132$&19111+1048&&G045.14+00.14&c1&4.53&Y\\
A17&J191641+102952&$45.140$&$-0.752$&19143+1024&&&&&\\
A18&J191602+103743&$45.183$&$-0.553$&&&G045.39-00.76&c6&7.18&Y\\
A19&J191539+104116&$45.190$&$-0.439$&19132+1035&&G045.09-00.51&c2&6.0&Y\\
A20&J191706+103032&$45.198$&$-0.838$&&&&&&\\
A21&J191633+104658&$45.378$&$-0.592$&19142+1041&&G045.74-00.26&c2&7.28&Y\\
&&&&&&G045.39-00.76&c3,c10&7.18&?\\
&&&&&&G045.84-00.56&c4&3.62&\\
A22&J191706+104322&$45.387$&$-0.738$&&&G045.39-00.76&c1,c9,c11,c13&7.18&Y\\
A23&J191701+104447&$45.399$&$-0.710$&&&G045.39-00.76&c7,c11,c13&7.18&Y\\
A24&J191620+105035&$45.407$&$-0.518$&19139+1045&&G045.84-00.56&c4&3.62&\\
A25*&J191451+110244&$45.416$&$-0.100$&19125+1057&&&&&\\
A26*&J191457+110324&$45.437$&$-0.116$&&&&&&\\
A27&J191532+105933&$45.447$&$-0.273$&&&&&&\\
A28&J191419+110902&$45.449$&$0.063$&19120+1103&&G045.49+00.04&c1,c5,c7,c8&7.45&Y\\
&&&&&&G045.14+00.14&c5&4.53&?\\
&&&&&&G045.24+00.19&c1&$-$&\\
A28a\tablenotemark{h}&J191414+110809&$45.425$&$0.077$&&&&&&\\
A29&J191425+110921&$45.464$&$0.046$&19120+1103\tablenotemark{i}&P3744&G045.49+00.04&c1,c4,c5,c7&7.45&Y\\
&&&&&&&c8&&\\
&&&&&&G045.14+00.14&c2&4.53&?\\
A30&J191407+111222&$45.475$&$0.133$&19117+1107&&G045.49+00.04&c2,c3,c5-c8&7.45&Y\\
&&&&&&G045.24+00.19&c1&$-$&\\
A30a&J191409+111245&$45.485$&$0.128$&&&&&&\\
A31&J191350+111548&$45.493$&$0.222$&&&G045.24+00.19&c1&$-$&\\
A32&J191536+110259&$45.506$&$-0.262$&&&&&&\\
A33&J191558+110113&$45.521$&$-0.354$&&&&&&\\
A34&J191608+110059&$45.536$&$-0.391$&19138+1055&&&&&\\
A35&J191450+111127&$45.543$&$-0.029$&&&G045.49+00.04&c4&7.45&Y\\
A36&J191446+111159&$45.544$&$-0.012$&19124+1106&&G045.49+00.04&c4&7.45&Y\\
A37&J191512+111007&$45.565$&$-0.119$&&&G045.49+00.04&c4&7.45&\\
&&&&&&G045.64-00.01&c2&11.12&Y\\
A38&J191459+111541&$45.623$&$-0.030$&&&G045.49+00.04&c4&7.45&\\
&&&&&&G045.64-00.01&c1,c3,c4&11.12&Y\\
A39&J191458+111639&$45.635$&$-0.018$&&&G045.49+00.04&c4&7.45&\\
&&&&&&G045.64-00.01&c1,c4&11.12&Y\\
A40&J191358+112532&$45.651$&$0.270$&&&G045.49+00.04&c9&7.45&\\
&&&&&&G045.89-00.36&c4&6.72&\\
&&&&&&G045.64+00.29&c1,c5,c6&1.88&\\
A41&J191709+110129&$45.660$&$-0.608$&&&G045.39-00.76&c2&7.18&Y\\
&&&&&&G045.94-00.56&c5&4.78&?\\
A42*&J191553+111347&$45.696$&$-0.237$&&&G046.19-00.56&c4&3.75&\\
&&&&&&G045.74-00.26&c1,c10,c11&7.28&Y\\
&&&&&&&c13-c15,c23,c24&&\\
&&&&&&G045.89-00.36&c2&6.72&?\\
&&&&&&G045.84-00.56&c3&3.62&\\
A43&J191705+110509&$45.707$&$-0.566$&&&&&&\\
A44&J191609+111423&$45.736$&$-0.291$&19137+1108&&G046.19-00.56&c4&3.75&\\
&&&&&&G045.74-00.26&c1,c8,c21&7.28&Y\\
&&&&&&G045.89-00.36&c2&6.72&\\
&&&&&&G046.14-00.21&c2&1.8&\\
&&&&&&G045.84-00.56&c3&3.62&\\
A45&J191604+111725&$45.772$&$-0.251$&&&G046.19-00.56&c4&3.75&\\
&&&&&&G045.74-00.26&c1,c12,c21&7.28&Y\\
&&&&&&G045.89-00.36&c2&6.72&?\\
&&&&&&G045.84-00.56&c3&3.62&\\
&&&&&&G046.09+00.24&c2&$-$&\\
A46&J191630+111606&$45.802$&$-0.355$&19141+1110&&G045.74-00.26&c16,c19&7.28&Y\\
&&&&&&G045.89-00.36&c2&6.72&?\\
&&&&&&G046.09+00.24&c2&$-$&\\
A47&J191618+111909&$45.823$&$-0.286$&19139+1113&&G046.19-00.56&c4&3.75&\\
&&&&&&G045.74-00.26&c1&7.28&?\\
&&&&&&G045.89-00.36&c2&6.72&Y\\
&&&&&&G045.84-00.56&c10&3.62&\\
&&&&&&G046.09+00.24&c2&$-$&\\
A48&J191713+111605&$45.884$&$-0.511$&&P3758&G045.89-00.36&c1&6.72&Y\\
A49&J191447+113705&$45.915$&$0.181$&19124+1131&&G045.89-00.36&c6&6.72&\\
&&&&&&G045.64+00.29&c1&1.88&\\
A50&J191655+112153&$45.935$&$-0.401$&19145+1116&&G045.89-00.36&c1,c2&6.72&Y\\
&&&&&&G045.94-00.56&c2&4.78&?\\
A51&J191514+113632&$45.959$&$0.078$&&&G045.89-00.36&c6&6.72&\\
&&&&&&G046.14-00.21&c6&1.8&\\
A52&J191456+114614&$46.068$&$0.218$&&&G045.89-00.36&c6&6.72&\\
&&&&&&G045.64+00.29&c1&1.88&\\
&&&&&&G046.09+00.24&c1&$-$&Y\\
A53&J191450+114805&$46.084$&$0.254$&19125+1142&&G045.89-00.36&c6&6.72&\\
&&&&&&G045.64+00.29&c1&1.88&\\
&&&&&&G046.09+00.24&c1&$-$&Y\\
A53a&J191448+114910&$46.096$&$0.270$&&&&&&\\
A54&J191413+115330&$46.094$&$0.429$&&&G046.14+00.39&c1&4.25&Y\\
&&&&&&G045.64+00.29&c1&1.88&\\
A55&J191423+115342&$46.115$&$0.396$&19120+1148&&G046.14+00.39&c1&4.25&Y\\
&&&&&&G045.64+00.29&c1&1.88&\\
A56&J191617+114240&$46.169$&$-0.102$&19139+1137&&G045.89-00.36&c6&6.72&\\
&&&&&&G046.14-00.21&c1&1.8&\\
A57&J191750+113102&$46.174$&$-0.527$&&P3762&G046.19-00.56&c1&3.75&Y\\
A58&J191400+120052&$46.177$&$0.535$&19116+1155&&&&&\\
A59&J191402+120157&$46.197$&$0.536$&&&&&&\\
A59a&&$46.212$&$ 0.551$&&&&&&\\
A60&J191819+113119&$46.233$&$-0.628$&&&G046.24-00.66&c1&4.07&Y\\
A61&J191851+112750&$46.244$&$-0.773$&&&&&&\\
A62&J191702+114640&$46.314$&$-0.234$&19146+1141&&G046.19-00.56&c2&3.75&?\\
&&&&&&G046.34-00.21&c2,c3,c5,c7&4.12&Y\\
&&&&&&&c9-c11&&\\
A63*&J191703+115012&$46.367$&$-0.208$&&&G046.19-00.56&c2&3.75&?\\
&&&&&&G046.34-00.21&c2,c5,c9,c12&4.12&Y\\
&&&&&&&c13&&\\
A64*&J191709+114956&$46.374$&$-0.231$&&&G046.19-00.56&c2&3.75&?\\
&&&&&&G046.34-00.21&c2,c5,c12&4.12&Y\\
A65&J191655+115407&$46.410$&$-0.150$&19146+1148&P3773&G046.19-00.56&c2&3.75&Y\\
&&&&&&G046.34-00.21&c2,c5,c12&4.12&\\
A66&J191715+115225&$46.423$&$-0.235$&19149+1147&P3776&G046.19-00.56&c2&3.75&?\\
&&&&&&G046.34-00.21&c1,c2,c5,c6&4.12&Y\\
&&&&&&&c12&&\\
A67*&J191711+115548&$46.465$&$-0.194$&&P3778&G046.19-00.56&c2&3.75&?\\
&&&&&&G046.34-00.21&c1,c5,c6,c12&4.12&Y\\
&&&&&&G046.79+00.04&c3&3.6&\\
A68&J191631+120108&$46.468$&$-0.009$&&&&&&\\
A69&J191630+120314&$46.496$&$0.013$&&&&&&\\
A70&J191849+115057&$46.581$&$-0.586$&19164+1145&&G046.19-00.56&c3&3.75&?\\
&&&&&&G046.89-00.36&c1&8.05&Y\\
A71&J191537+121816&$46.618$&$0.319$&19132+1212&&G046.59+00.34&c1&0.65&Y\\
&&&&&&G046.59+00.19&c2&0.8&?\\
\hline \hline \\
\multicolumn{10}{l}{{$^a$ Source names include the prefix `BLAST' as part of the standard BLAST nomenclature.}}\\
\multicolumn{10}{l}{{$^b$ Closest \IRAS\ match within 30\arcsec\ in the Point Source Catalog (PSC).}}\\
\multicolumn{10}{l}{{$^c$ Closest IRDC match within 30\arcsec\ in the catalog from \citet{peretto2009}.}}\\
\multicolumn{10}{l}{{$^d$ GRS name of the extended clouds from \citet{rathborne2009} which contain the molecular clump}}\\ 
\multicolumn{10}{l}{{counterpart(s) to the BLAST source. A blank implies the absence of a clump counterpart in the catalog.}}\\
\multicolumn{10}{l}{{$^e$ Clump(s) from \citet{rathborne2009} containing the BLAST peak within its/their $\ell$ and $b$ FWHM.}}\\
\multicolumn{10}{l}{{$^f$ Distance from \citet{roman2009}.}}\\
\multicolumn{10}{l}{{$^g$ Cloud association: (Y) - BLAST clump is associated with the cloud. Cloud has a molecular clump whose}}\\
\multicolumn{10}{l}{{ velocity FWHM contains the strongest molecular peak of spectrum, measured at the peak of the submm emission.}}\\
\multicolumn{10}{l}{{ Molecular clump also has peak velocity closest to that measured for the BLAST source.}}\\
\multicolumn{10}{l}{{ (?) - Cloud membership unclear. The main molecular emission in the BLAST spectrum is within the velocity}}\\
\multicolumn{10}{l}{{ FWHM of this molecular cloud/clump, but there is still another cloud with a clump with peak velocity}}\\
\multicolumn{10}{l}{{ closer to that of the strongest line in the BLAST spectrum. Contributions from different clouds are possible.}}\\
\multicolumn{10}{l}{{ A blank implies that the velocity of main BLAST molecular emission is not compatible with that assigned }}\\
\multicolumn{10}{l}{{ to the cloud/clump, despite the spatial agreement and apparent location of the submm peak within the}}\\
\multicolumn{10}{l}{{ spatial FWHM of clump(s) associated with that cloud.}}\\
\multicolumn{10}{l}{{$^h$ Source designations ending in `a' refer to additional Gaussians required during the fit of the main clump.}}\\
\multicolumn{10}{l}{{$^i$ Position beyond 30\arcsec, but visually within the \IRAS\ emission.}}\\
\multicolumn{10}{l}{{* Detection S/N$<3$.}}
\end{longtable*}

\section{Discussion of Individual Sources} \label{sec:sources}
In the following sections we apply the results of our analysis to describe the main submm regions (G045.49+00.04, G045.74-00.26, G045.89-00.36, and G045.14+00.14), adopting the designations from Rathborne et al (2009; Figure \ref{fig:molecular}).
\subsection{GRSMC G045.14+00.14}
\subsubsection{G45.12}
\begin{figure}[t]
\includegraphics[scale=0.48,angle=270]{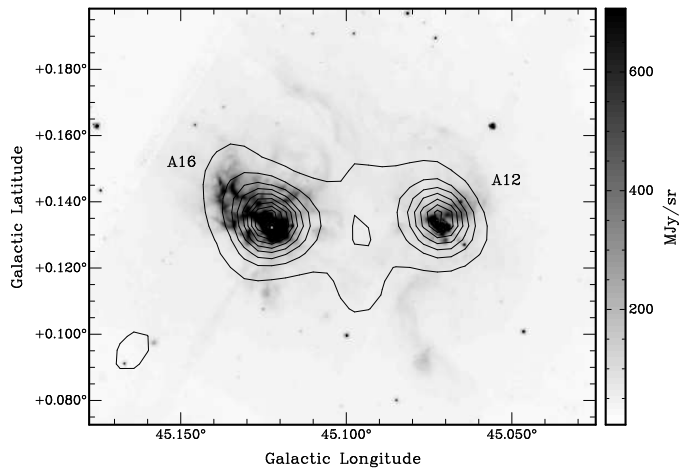}
\caption{Greyscale \irac\ 8\,\micron\ image of IRAS 19111+1048 (A16) and IRAS 19110+1045 (A12) with BLAST 500\,\micron\ contours superimposed. Contours are from 10 to 90\% of map peak value 2153\,MJy\,sr$^{-1}$ in 10\% steps.}
\label{fig:irac1}
\end{figure}

GRSMC 45.122+0.132 (G45.12; IRAS 19111+1048; A16 in the present work), contains a cometary UCH{\sc ii}R (WC89) classified as a massive young stellar object (MYSO) by \citet{chan1996}. 
It shows extended submm emission towards the north (upper left of Figure \ref{fig:irac1}), observed in the MIR as a bright rimmed \ion{H}{2} region. This structure is likely produced by the activity of the young massive population embedded in the innermost regions of the clump, as detected by \citet{vig2006} using the Giant Metrewave Radio Telescope (GMRT) at 1200 and 610\,MHz. The 19 embedded OB stars observed by these authors are located within and surrounding the peak of our BLAST clump, mainly within the \ion{H}{2} region. Their brightest source agrees well with our submm peak ($\lesssim5$\arcsec), and although it remained unresolved in their datasets, they suspected it to be a cluster responsible for the extended morphology in the radio. A possible cluster powering this region has also been suggested through [NeII] observations by \citet{takahashi2000}, and later by \citet{zhu2008}, who detected three additional peaks (G45.12N, G45.12SE and G45.12SW) nearby the main UCH{\sc ii}R from WC89.

\begin{table*}[t!]
\caption{Coordinates of CORNISH sources in the BLAST Aquila field.}
\label{table:rsrcs}
\centering
\begin{tabular}{l l l l | l l l l}
\hline
\hline
BLAST&CORNISH&RA&Dec&BLAST&CORNISH&RA&Dec\\
&&(J2000)&(J2000)&&&(J2000)&(J2000)\\
&&h m s&\degree\ \arcmin\ \arcsec&&&h m s&\degree\ \arcmin\ \arcsec\\
\hline
A12&A12$-$C0&19 13 19.063&+10 51 26.521&A28&A28$-$C0&19 14 20.710&+11 09 13.540\\
&A12$-$C1*&19 13 21.617&+10 50 56.756&&A28$-$C1&19 14 20.714&+11 08 57.156\\
&A12$-$C2&19 13 21.886&+10 50 49.704&&A28$-$C2&19 14 20.784&+11 09 17.737\\
&A12$-$C3&19 13 22.130&+10 50 52.163&&A28$-$C3&19 14 20.918&+11 09 15.926\\
&A12$-$C4&19 13 23.834&+10 51 42.624&&A28$-$C4&19 14 20.957&+11 09 01.728\\
&A12$-$C5&19 13 23.954&+10 51 41.497&&A28$-$C5&19 14 21.002&+11 09 16.578\\
&A12$-$C6&19 13 25.548&+10 51 07.567&&A28$-$C6&19 14 21.002&+11 09 20.246\\
&A12$-$C7&19 13 25.584&+10 51 07.535&&A28$-$C7&19 14 21.120&+11 09 08.550\\
&A12$-$C8&19 13 25.601&+10 51 09.320&&A28$-$C8&19 14 21.190&+11 09 07.268\\
&A12$-$C9&19 13 25.697&+10 50 57.977&&A28$-$C9&19 14 21.226&+11 09 02.110\\
A16&A16$-$C0&19 13 24.283&+10 53 28.525&&A28$-$C10&19 14 21.250&+11 09 14.915\\
&A16$-$C1*&19 13 24.374&+10 53 27.247&&A28$-$C11&19 14 21.276&+11 09 11.534\\
&A16$-$C2&19 13 24.446&+10 53 25.919&&A28$-$C12&19 14 21.338&+11 09 22.162\\
&A16$-$C3&19 13 24.994&+10 53 13.895&&A28$-$C13&19 14 21.362&+11 09 10.084\\
&A16$-$C4&19 13 25.116&+10 54 16.153&&A28$-$C14&19 14 21.386&+11 09 11.242\\
&A16$-$C5&19 13 25.205&+10 53 14.932&&A28$-$C15&19 14 21.408&+11 09 13.309\\
&A16$-$C6&19 13 27.055&+10 53 19.946&&A28$-$C16&19 14 21.578&+11 09 04.824\\
&A16$-$C7&19 13 27.218&+10 53 22.794&&A28$-$C17&19 14 21.617&+11 09 15.264\\
&A16$-$C8&19 13 27.331&+10 53 46.489&&A28$-$C18&19 14 21.720&+11 08 55.424\\
&A16$-$C9*&19 13 27.516&+10 53 46.558&&A28$-$C19&19 14 21.854&+11 08 54.413\\
&A16$-$C10&19 13 27.545&+10 53 30.196&&A28$-$C20&19 14 21.862&+11 09 30.956\\
&A16$-$C11*&19 13 27.583&+10 53 46.237&&A28$-$C21&19 14 21.893&+11 09 17.791\\
&A16$-$C12*&19 13 27.583&+10 53 49.636&&A28$-$C22&19 14 22.022&+11 09 45.385\\
&A16$-$C13&19 13 27.586&+10 53 27.920&&A28$-$C23&19 14 22.099&+11 09 45.972\\
&A16$-$C14&19 13 27.602&+10 53 25.483&&A28$-$C24&19 14 22.272&+11 09 37.760\\
&A16$-$C15&19 13 27.703&+10 53 29.249&&A28$-$C25&19 14 22.481&+11 09 31.090\\
&A16$-$C16&19 13 27.710&+10 53 29.807&&A28$-$C26&19 14 22.615&+11 09 43.690\\
&A16$-$C17&19 13 27.763&+10 53 38.990&&A28$-$C27&19 14 22.776&+11 08 57.970\\
&A16$-$C18&19 13 27.773&+10 53 34.897&A29&A29$-$C0&19 14 25.690&+11 09 25.330\\
&A16$-$C19&19 13 27.797&+10 53 32.194&&A29$-$C1&19 14 26.222&+11 08 35.488\\
&A16$-$C20&19 13 27.866&+10 53 30.181&&A29$-$C2&19 14 26.942&+11 09 14.897\\
&A16$-$C21&19 13 27.888&+10 53 40.402&&A29$-$C3&19 14 27.079&+11 09 12.456\\
&A16$-$C22&19 13 27.912&+10 53 33.659&&A29$-$C4&19 14 27.559&+11 09 56.617\\
&A16$-$C23&19 13 27.929&+10 53 27.013&A30&A30$-$C0&19 14 07.330&+11 12 41.742\\
&A16$-$C24&19 13 27.965&+10 53 35.675&&A30$-$C1*&19 14 08.227&+11 12 22.684\\
&A16$-$C25&19 13 28.008&+10 53 45.150&&A30$-$C2&19 14 08.232&+11 12 36.929\\
&A16$-$C26&19 13 28.022&+10 53 28.806&&A30$-$C3&19 14 08.602&+11 12 26.327\\
&A16$-$C27&19 13 28.030&+10 53 24.598&&A30$-$C4*&19 14 08.830&+11 12 26.586\\
&A16$-$C28&19 13 28.034&+10 53 42.302&&A30$-$C5&19 14 09.046&+11 12 25.661\\
&A16$-$C29&19 13 28.128&+10 53 31.708&&A30$-$C6&19 14 09.094&+11 12 22.075\\
&A16$-$C30&19 13 28.212&+10 53 37.547&A30a&A30a$-$C0&19 14 08.604&+11 13 19.175\\
&A16$-$C31&19 13 28.222&+10 53 34.260&A47&A47$-$C0&19 16 17.645&+11 18 59.670\\
&A16$-$C32&19 13 28.226&+10 53 35.905&&A47$-$C1&19 16 17.746&+11 19 14.509\\
&A16$-$C33&19 13 28.380&+10 53 32.298&&A47$-$C2&19 16 19.330&+11 19 14.992\\
&A16$-$C34&19 13 29.842&+10 53 34.318&A50&A50$-$C0&19 16 55.730&+11 21 48.355\\
&A16$-$C35&19 13 31.090&+10 53 45.031&&&&\\
\hline
\multicolumn{8}{l}{{* Detection S/N$<3$.}}\\
\end{tabular}
\end{table*}

\begin{table*}[ht!]
\caption{Spectral types and IR counterparts for the BLAST peaks A16 and A12 and the radio sources detected in the CORNISH data within $\sim1$\arcmin\ of the submm peaks.}
\label{table:radio}
\centering
\begin{tabular}{l l l l l l}
\hline
\hline
Source&IR+submm$^a$&IR+submm$^a$&IR$^b$&TFT1$^c$&TFT2$^c$\\
&(ZAMS)&(Class V)&&(6\,cm)&(3.6\,cm)\\
\hline
\hline
&log$Q_0$$^d$&4.8\,GHz:ZAMS/Class V&IR$^e$&3.6\,cm$^e$&GLIMPSE$^f$\\
&[s$^{-1}$]&&&&\\
\hline
\hline
A16&O6$-$O5.5&O6$-$O4&O5&O6&O6\\
A16$-$C0&$46.86^{+0.01}_{-0.01}$&B0.5$-$B0/$-$&&&\\
A16$-$C1&$46.96^{+0.16}_{-0.26}$&B0.5$-$B0/$-$&&&\\
A16$-$C2&$46.97^{+0.01}_{-0.01}$&B0/$-$&&&\\
A16$-$C3&$47.36^{+0.00}_{-0.00}$&B0/$-$&&&\\
A16$-$C4&$46.70^{+0.01}_{-0.01}$&B0.5/$-$&&&\\
A16$-$C5&$46.83^{+0.01}_{-0.01}$&B0.5$-$B0/$-$&&&\\
A16$-$C6&$47.06^{+0.17}_{-0.27}$&B0.5$-$B0/$-$&&&G045.1163+00.1327\\
A16$-$C7&$47.03^{+0.01}_{-0.01}$&B0/$-$&&&G045.1163+00.1327\\
A16$-$C8&$46.68^{+0.04}_{-0.05}$&B0.5/$-$&&&\\
A16$-$C9&$46.71^{+0.05}_{-0.06}$&B0.5/$-$&&&\\
A16$-$C10&$47.58^{+0.01}_{-0.01}$&B0/$-$&&&\\
A16$-$C11&$46.51^{+0.01}_{-0.01}$&B0.5/$-$&&&\\
A16$-$C12&$46.73^{+0.01}_{-0.01}$&B0.5/$-$&&&\\
A16$-$C13&$47.33^{+0.12}_{-0.16}$&B0/$-$&&&\\
A16$-$C14&$47.73^{+0.08}_{-0.09}$&O9.5/$-$&&&\\
A16$-$C15&$47.54^{+0.05}_{-0.05}$&B0/$-$&&&\\
A16$-$C16&$47.63^{+0.16}_{-0.25}$&B0$-$O9.5/$-$&&&\\
A16$-$C17&$48.30^{+0.03}_{-0.03}$&O8.5$-$O8/O8.5&&&G045.1221+00.1322\\
A16$-$C18$^g$&$47.74^{+0.44}_{-47.74}$&$<$O8.5/O8.5&&&G045.1221+00.1322\\
A16$-$C19&$47.89^{+0.01}_{-0.01}$&O9.5/O9.5&&&G045.1221+00.1322\\
A16$-$C20&$47.79^{+0.01}_{-0.01}$&O9.5/$-$&&&\\
A16$-$C21&$48.25^{+0.20}_{-0.38}$&O9.5$-$O7.5/O8&&&G045.1221+00.1322\\
A16$-$C22$^g$&$48.16^{+0.01}_{-0.01}$&O8.5/O9$-$O8.5&&&G045.1221+00.1322\\
A16$-$C23&$46.88^{+0.02}_{-0.02}$&B0.5$-$B0/$-$&&&\\
A16$-$C24&$49.14^{+0.00}_{-0.00}$&O6/O5.5&O5&O6&G045.1221+00.1322\\
A16$-$C25&$47.53^{+0.07}_{-0.08}$&B0/$-$&&&\\
A16$-$C26&$47.90^{+0.05}_{-0.06}$&O9.5/O9.5&B0.5&B0&\\
A16$-$C27&$46.59^{+0.04}_{-0.05}$&B0.5/$-$&&&\\
A16$-$C28&$47.98^{+0.02}_{-0.02}$&O9.5$-$O9/O9.5$-$O9&&&\\
A16$-$C29&$47.87^{+0.01}_{-0.01}$&O9.5/$-$&&&\\
A16$-$C30&$48.09^{+0.16}_{-0.26}$&O9.5$-$O8.5/O8.5&&&\\
A16$-$C31&$47.93^{+0.23}_{-0.52}$&B0$-$O8.5/O9&&&\\
A16$-$C32&$48.29^{+0.01}_{-0.01}$&O8/O8.5&&&\\
A16$-$C33&$47.58^{+0.29}_{-1.18}$&B0.5$-$09.5/$-$&&&\\
A16$-$C34&$47.06^{+0.01}_{-0.01}$&B0/$-$&&&\\
A16$-$C35&$47.65^{+0.00}_{-0.00}$&O9.5/$-$&&&G045.1302+00.1223\\
\hline
A12&O6&O5&O6&O8.5$^h$&O8$^h$\\
A12$-$C0&$47.32^{+0.04}_{-0.05}$&B0/$-$&&&G045.0729+00.1484\\
A12$-$C1&$46.58^{+0.00}_{-0.00}$&B0.5/$-$&&&\\
A12$-$C2&$46.49^{+0.00}_{-0.00}$&B0.5/$-$&B0.5&B0$-$O9.5&G045.0712+00.1321\\
A12$-$C3&$47.88^{+0.00}_{-0.00}$&O9.5/$-$&O6.5&O9&G045.0712+00.1321\\
A12$-$C4&$47.53^{+0.07}_{-0.08}$&B0$-$O9.5/$-$&&&\\
A12$-$C5&$47.99^{+0.00}_{-0.00}$&O9/O9&&&\\
A12$-$C6&$46.47^{+0.02}_{-0.02}$&B0.5/$-$&&&\\
A12$-$C7$^g$&$48.15^{+0.2}_{-0.39}$&O9.5$-$O9/$<$O8.5&&&\\
A12$-$C8&$47.03^{+0.09}_{-0.11}$&B0/$-$&&&\\
A12$-$C9&$47.32^{+0.03}_{-0.03}$&B0/$-$&&&G045.0788+00.1191\\
\hline
\multicolumn{6}{l}{{$^a$ ZAMS and Class V spectral types from \citet{panagia1973} and \citet{martins2005} for}}\\
\multicolumn{6}{l}{{$D=7$\,kpc. If the luminosity is not within the range of the calibration scale, no estimate ($-$) is provided.}}\\
\multicolumn{6}{l}{{$^b$ ZAMS spectral type derived by \citet{kraemer2003} from the total source flux in the IR (\IRAS) for $D=6$\,kpc.}}\\
\multicolumn{6}{l}{{$^c$ ZAMS spectral type of the counterpart of the brightest CORNISH source within BLAST clump}}\\
\multicolumn{6}{l}{{in \citet{testi1999} for $D=7$\,kpc.}}\\
\multicolumn{6}{l}{{$^d$ Total number of ionizing photons per unit time for a distance of 7\,kpc and $T_{\mathrm{e}}\sim8000$\,K, assuming}}\\
\multicolumn{6}{l}{{optically thin emission.}}\\
\multicolumn{6}{l}{{$^e$ ZAMS spectral type derived by \citet{kraemer2003} for $D=6$\,kpc. Values are provided only for sources lying closer}}\\
\multicolumn{6}{l}{{than 5\arcsec\ to the CORNISH source.}}\\
\multicolumn{6}{l}{{$^f$ Closest IRAC GLIMPSE I Complete Catalog match within $\sim5$\arcsec\ of the CORNISH peak.}}\\
\multicolumn{6}{l}{{$^g$ Extra source required during Gaussian fitting.}}\\
\multicolumn{6}{l}{{$^h$ Estimate includes total flux of C0 and C1, as both are unresolved at 6\,cm \citep{testi1999}.}}\\
\end{tabular}
\end{table*}

\begin{figure}[ht]
\includegraphics[scale=0.47,angle=270]{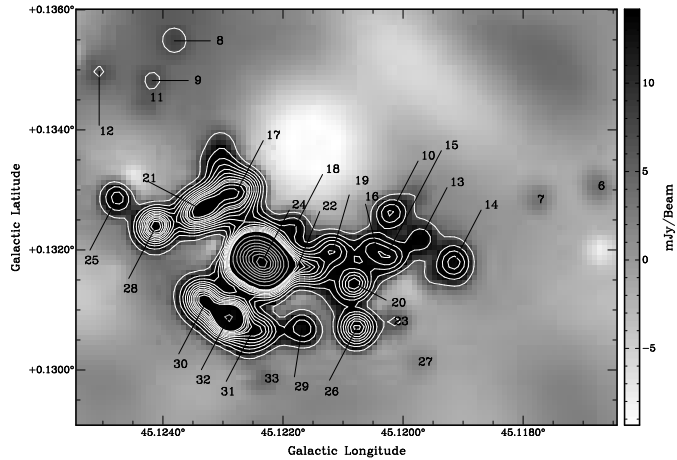}
\caption{Greyscale CORNISH 4.8\,GHz image of central cluster (S14 in \citealp{vig2006}) in IRAS 19111+1048 (A16) with numbered emission peaks (Table \ref{table:radio}). White contours are from 2.5\% to 12.5\% of map peak value 0.7\,Jy\,beam$^{-1}$ in 1\% steps. Grey contours are from 15\% to 85\% in 10\% steps.}
\label{fig:cornish1a}
\end{figure}

Our analysis of the 4.8\,GHz CORNISH data confirms such suspicions, with the detection of 28 radio sources in a region $\sim0.8$\,pc in radius centered on (and including) the central brightest member (A16$-$C24; Table \ref{table:radio}; Figure \ref{fig:cornish1a}). These objects are believed to be forming the unresolved source observed by Vig et al. (2006, S14 in their sample). None have a radio ZAMS spectral type later than B0 (with the exception of A16$-$C18, which was not initially detected by our extraction routine). 

We find that 22 of these radio sources lie within $\sim0.5$\,pc of A16$-$C24. This object has a bolometric luminosity equivalent to an O6 ZAMS star, which is in good agreement with previous results for a distance of $\sim6$\,kpc (e.g., \citealp{kraemer2003}). The position of A16$-$C24 also agrees with the UCH{\sc ii}R of WC89 and \citet{testi1999}. The two 2MASS IR sources detected by \citet{vig2006} within their S14 source (IR4 and IR5) agree well with A16$-$C24 and A16$-$C26 and the two peaks (KJK1 and KJK2) detected by \citet{kraemer2003}. The G45.12SE peak from \citet{zhu2008} traces the region of A16$-$C26; G45.12SW is formed by the conglomeration of radio sources towards A16$-$C13, and G45.12N is centered on the arc-shaped radio emission around A16$-$C21. The estimated spectral types of all radio sources within $\sim1$\arcmin\ of the BLAST coordinates have been included in Table \ref{table:radio} for an electron temperature similar to that estimated by Vig et al. (2006, $\sim8000$\,K, for their source containing the cluster: S14).

Taking into account the 3\,$\sigma$ positional error of our CORNISH source A16$-$C24 ($\sim0.5$\arcsec) and that expected for the OH maser detection of \citet{baart1985}, $\sim10$ milliarcsec, the separation of $\sim3$\arcsec\ between the two may be physically significant. 
Indeed, our projected separation of $\sim0.1$\,pc positions the maser source at the edge of the UCH{\sc ii}R, and agrees well with the displacement measured by \citet{baart1985}, of the order of $\sim0.03$\,pc (with respect to the coordinates from \citealp{matthews1977}). This distance, as pointed out by \citet{baart1985}, is well within the upper limit for the radius at which OH masers are produced in a shell-like front surrounding a massive young star. This suggests that these may be dense fragments of the expanding shell. The blueshift for the maser clusters detected by these same authors ($\sim6$\,km\,s$^{-1}$) is also similar to that observed for the dense gas traced with ammonia by \citet{hofner1999}, whose absorption peak coordinates are also offset from our radio peak in the same direction as the maser source. This supports the real physical origin of the observed displacement.

\begin{figure}[ht]
\includegraphics[scale=0.5,angle=270]{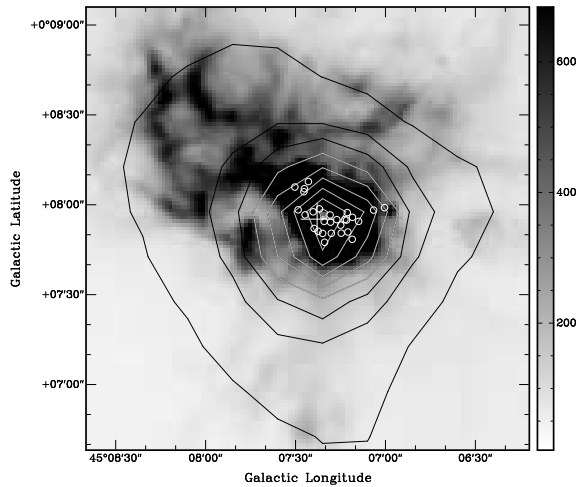}
\caption{Greyscale \irac\ 8\,\micron\ image of IRAS 19111+1048 (A16) with 250\,\micron\ BLAST contours overlaid. Cross marks the position of submm peak, and circles the positions of main CORNISH detections (central cluster). Contours are from 5\% to 85\% of the map peak value of 31500\,MJy\,sr$^{-1}$ in 10\% steps.}
\label{fig:cornish1a_b}
\end{figure}

Although further investigation is required, our analysis of the CORNISH data and the evidence in the mid infrared (MIR) and submm suggest that A16$-$C24 may be the initial trigger and power source of G45.12. The massive outflow detected by \citet{hunter1997} appears to follow the overall direction of the distribution of CORNISH sources well, which could indicate that A16$-$C24 has induced the formation of its lower mass companions. In addition, the distribution of CORNISH sources appears to have a slightly curved, bow-shock like morphology pointing towards the submm comma-like structure extending downwards in Figure \ref{fig:cornish1a_b}. This structure also coincides with the strongest emission in $^{13}$CO and CS. Considering its central position, A16$-$C24 appears as a suitable candidate for the main triggering/compressing source. 

\subsubsection{G45.07}
Just like its neighbor, IRAS 19110+1045 (A12; GRSMC 45.073+0.129 [G45.07]) also shows submm, IR, and radio structure signaling the presence of massive star formation, albeit significantly less pronounced than in G45.12. Only nine possible CORNISH detections are found within $\sim1$\arcmin\ of the BLAST position, with radio ZAMS/class V spectral types ranging between B0.5 and O9 for an electron temperature of 8000\,K (Table \ref{table:radio}). The lower number of stars and their later apparent radio spectral type is consistent with the more compact size of the resulting ionized region compared to G45.12. 

\begin{figure}[ht]
\includegraphics[scale=0.5,angle=270]{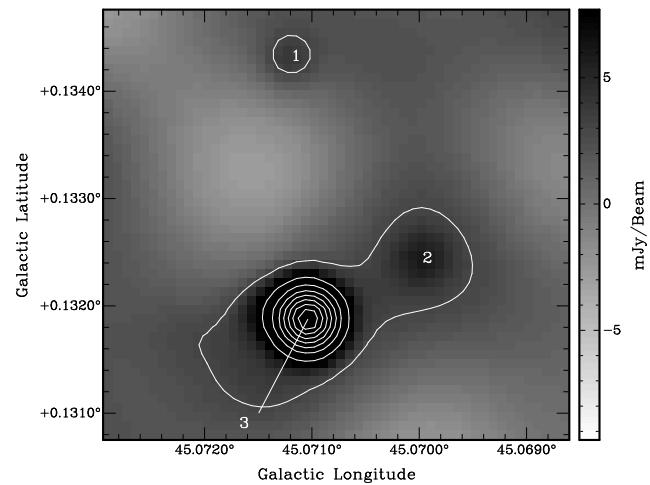}
\caption{Greyscale CORNISH image of three central sources likely powering IRAS 19110+1045 (A12). Contours are from 2\% to 9\% of map peak value 0.7\,Jy\,beam$^{-1}$ in 1\% steps.}
\label{fig:cornish1b}
\end{figure}

Two CORNISH sources, A12$-$C2 and A12$-$C3 (Figure \ref{fig:cornish1b}), match ($\lesssim3$\arcsec.5) those detected by \citet{vig2006} and the two embedded IR sources (KJK1, KJK3) from \citet{kraemer2003}. Despite the positional accuracy of our CORNISH measurements ($<1$\arcsec.5 difference between our two coordinates and those given by \citealp{testi1999} and WC89), we find no indication of the third IR source (KJK2) detected by \citet{kraemer2003}. This source is located $\sim2$\arcsec\ from A12$-$C3, and should have been identifiable at the resolution limit of the CORNISH data.

\begin{figure}[ht]
\includegraphics[scale=0.5,angle=270]{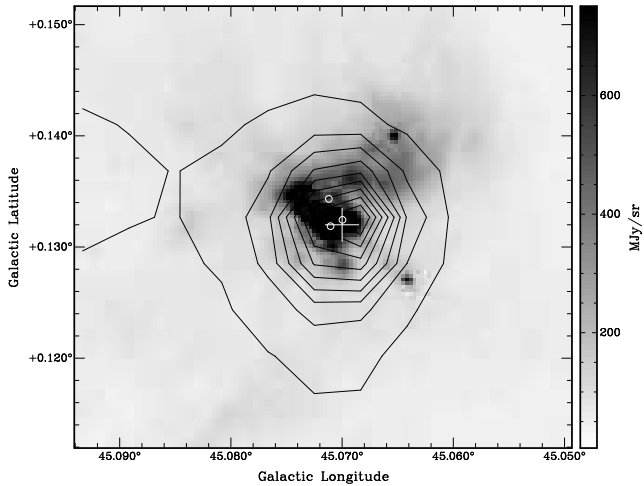}
\caption{Greyscale \irac\ 8\,\micron\ image of IRAS 19110+1045 (A12). Symbols and contours like Fig.\ref{fig:cornish1a_b}.}
\label{fig:cornish1b_b}
\end{figure}

As with G45.12, G45.07 also contains a UCH{\sc ii}R (spherical/unresolved UCH{\sc ii}R; WC89) with a CO outflow \citep{hunter1997} and masers. Our brightest radio source lies at $<1$\arcsec.5 from the coordinates given by WC89. The masers show a more significant displacement ($\sim2$\arcsec), and appear to extend along the overall outflow direction. In addition to hydroxyl masers, G45.07 also shows several water maser sources (e.g., \citealp{hofner1996}) and 6.7\,GHz methanol emission \citep{pandian2007a}, with distances ranging between $\sim1$-3\arcsec.5 and $\sim10$\arcsec.5 from our brightest CORNISH radio peak, respectively.

Most water masers align well with our radio source A12$-$C3 and the overall outflow direction. This, together with the clear separation from our radio peak, again supports the scenario where they are energized by the bipolar outflow (e.g., \citealp{hofner1996}; \citealp{buizer2005}). The presence of these water masers also implies a less advanced evolutionary stage than that of its neighbor G45.12 (in agreement with previous studies: e.g., \citealp{vig2006}). This is supported by the simpler overall morphology, the lower stellar content, and the much earlier spectral type inferred from the IR/submm for A12$-$C3 and for the total clump with respect to those derived from the radio, suggesting a much more deeply embedded stage with significant radio optical depth.

While there appears to be extended MIR emission in the GLIMPSE images, consistent with the outflow orientation, there is also considerable emission in an almost orthogonal direction to its main axis  (western upper right corner in Figure \ref{fig:cornish1b_b}). The peak of the 21\,cm radio continuum is displaced towards this region. The overall morphology appears to be similar to the extended \ion{H}{2} in G45.12, albeit in a smaller scale, which could be due to its younger age and/or the lower stellar content in the innermost central region.

\begin{table*}[ht!]
\caption{Spectral types and IR counterparts for the BLAST peaks A28, A29, A30, and A30a and the radio sources detected in the CORNISH data within $\sim1$\arcmin\ of the submm peaks.}
\label{table:radio1}
\centering
\begin{tabular}{l l l l l l}
\hline
\hline
Source&IR+submm&IR+submm&IR&TFT1&TFT2\\
&(ZAMS)&(Class V)&&(6\,cm)&(3.6\,cm)\\
\hline
\hline
&log$Q_0$&4.8\,GHz:ZAMS/Class V&IR&3.6\,cm&GLIMPSE\\
&[s$^{-1}$]&&&&\\
\hline
\hline
AA29&O6.5$-$O5.5&O7$-$O4.5&&&\\
A28&&&O5.5&&O6.5\tablenotemark{a}$-$O5.5\tablenotemark{b}\\
Complex1&$49.25^{+0.00}_{-0.00}$&O5.5/O5&&&G045.4509+00.0570\tablenotemark{c}\\
A28$-$C1&$47.35^{+0.04}_{-0.04}$&B0/$-$&&&\\
A28$-$C18&$46.79^{+0.15}_{-0.23}$&B0.5$-$B0/$-$&&&\\
A28$-$C19&$47.19^{+0.01}_{-0.01}$&B0/$-$&&&\\
A28$-$C20&$47.11^{+0.11}_{-0.16}$&B0/$-$&&&\\
A28$-$C22&$46.38^{+0.21}_{-0.42}$&B0.5/$-$&&&\\
A28$-$C23&$47.13^{+0.15}_{-0.23}$&B0.5$-$B0/$-$&&&\\
A28$-$C24&$46.69^{+0.03}_{-0.04}$&B0.5/$-$&&&\\
A28$-$C25&$46.62^{+0.02}_{-0.02}$&B0.5/$-$&&&\\
A28$-$C26&$46.91^{+0.02}_{-0.02}$&B0.5$-$B0/$-$&&&G045.4640+00.0574\\
A28$-$C27&$46.69^{+0.02}_{-0.02}$&B0.5/$-$&&&\\
A29&&&&&O9.5\\
A29$-$C0&$47.84^{+0.00}_{-0.00}$&O9.5/$-$&&&G045.4661+00.0457(GR)\\
&&&&&G045.4658+00.0452 (GC)\\
A29$-$C1&$46.44^{+0.02}_{-0.02}$&B0.5$-$B0/$-$&&&G045.4551+00.0359\\
A29$-$C2&$47.17^{+0.01}_{-0.01}$&B0/$-$&&&\\
A29$-$C3&$47.28^{+0.00}_{-0.01}$&B0/$-$&&&\\
A29$-$C4&$47.29^{+0.02}_{-0.02}$&B0/$-$&&&\\
\hline
AA30&B0$-$O6.5&$<$O6.5&&&\\
A30&&&O6.5$-$O6&&O6.5\\
Complex2&$48.47^{+0.00}_{-0.00}$&O7.5/O8&O6.5&O7&\\
A30$-$C0&$46.78^{+0.04}_{-0.04}$&B0.5$-$B0/$-$&&&G045.4790+00.1365\\
A30$-$C1&$47.09^{+0.06}_{-0.06}$&B0/$-$&&&\\
A30$-$C2&$47.27^{+0.11}_{-0.16}$&B0/$-$&&&\\
A30a&&&&&\\
A30a$-$C0&$46.49^{+0.03}_{-0.03}$&B0.5/$-$&&&G045.4909+00.1376(GR)\\
&&&&&G045.4904+00.1367(GC)\\
\hline
A46&B0.5$-$B0&&&&\\
\hline
A47&O8$-$O7.5&O9$-$O8&&&\\
A47$-$C0&46.93$^{+0.26}_{-0.74}$&B0.5$-$B0/$-$&&&G045.8202-00.2866\\
A47$-$C1&46.61$^{+0.06}_{-0.08}$&B0.5/$-$&&&G045.8232-00.2842 (GR)\\
A47$-$C2&46.73$^{+0.08}_{-0.09}$&B0.5$-$B0/$-$&&&G045.8266-00.2892 (GR)\\
\hline
A50&O9.5$-$O8.5&$<$O9.5&&&\\
A50$-$C0&46.97$^{+0.09}_{-0.11}$&B0.5$-$B0/$-$&&&G045.9344-00.4016\\
\hline
\multicolumn{6}{l}{{$^a$ Compact source only, $\sim6$\arcsec\ in size, and therefore smaller than our aperture used for}}\\
\multicolumn{6}{l}{{photometry.}}\\
\multicolumn{6}{l}{{$^b$ Compact source including the extended emission.}}\\
\multicolumn{6}{l}{{$^c$ The closest CORNISH counterpart to this source is C54 (Table \ref{table:rsrcs}).}}\\
\end{tabular}
\end{table*}

\subsubsection{Submm Analysis}\label{submm1}
Our analysis of the SEDs yields total system masses for G45.12 and G45.07 of $\sim 3550$\,M$_\odot$ and $\sim1400$\,M$_\odot$, respectively, for a distance of 7\,kpc (Table \ref{table:seds}). 
This differs from the estimates given by \citet{hunter1997}, who obtained (with temperature, dust emissivity index, and optical depth as free parameters)  masses $\sim7$ and $\sim4$ times larger than our estimated masses, respectively (when scaled to their distance of 8.3\,kpc). \citet{vig2006} obtained, through SED fitting and radiative transfer modeling (assuming a spherically symmetric homogeneous cloud of gas and dust and a distance of 6\,kpc), masses of 5000\,M$_\odot$ ($\sim20$ M$_\odot$ of dust, and a larger gas-to-dust ratio of 250) and 13,000\,M$_\odot$ (for a dust mass of 30\,M$_\odot$, and gas to dust ratio of 450) for G45.12 and G45.07, respectively. Scaling to the same distance, we obtain $M_{\mathrm{d}}\sim26$ and $M_{\mathrm{d}}\sim10.5$\,M$_\odot$. While our dust mass estimate for G45.12 is in good agreement with their estimate from the SCUBA maps, the mass for G45.07 is in better agreement with the mass these authors estimated from their 130\,\micron\ map (obtained with the TIFR 1-m balloon borne telescope; \citealp{ghosh1988}), of $\sim13$\,M$_\odot$.

We find temperatures for G45.12 and G45.07 of $\sim40$\,K, and clump luminosities equivalent to an $\sim$O6 ZAMS spectral type star  ($\sim$O5 Class V star). This is in good agreement with those estimated from the IR (e.g., \citealp{kraemer2003}) for the brightest sources within the clumps. This suggests that A16$-$C24 and A12$-$C3 are the main sources responsible for the dust heating in G45.12 and G45.07, respectively. We note, however, that the significantly later spectral type derived from the radio for A12$-$C3 appears to suggest a much younger and highly embedded stage than A16$-$C24.

\subsection{GRSMC G045.49+00.04}
\begin{figure}[ht]
\includegraphics[scale=0.48,angle=270]{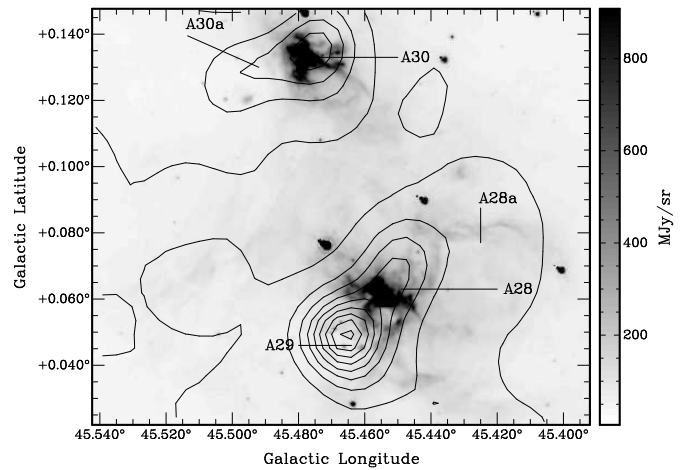}
\caption{Greyscale \irac\ 8\,\micron\ image of IRAS 19120+1103 and IRAS 19117+1107. Contours like Fig.\ref{fig:irac1}.}
\label{fig:irac2}
\end{figure}

\begin{figure}[ht]
\includegraphics[scale=0.50,angle=270]{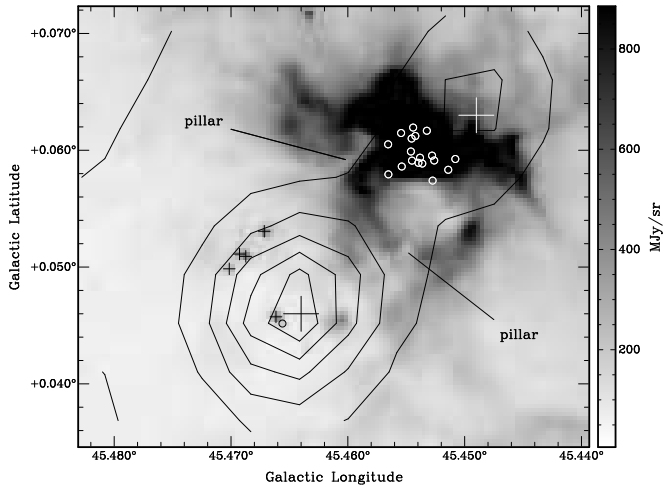}
\caption{Greyscale \irac\ 8\,\micron\ image of IRAS 19120+1103 (A28$-$A29) with 250\,\micron\ BLAST contours overlaid. Crosses mark the position of main GR sources. Contours are from 5\% to 55\% of the map peak value of 31500\,MJy\,sr$^{-1}$ in 10\% steps. Other symbols like Fig.\ref{fig:cornish1a_b}.}
\label{fig:BB70irac}
\end{figure}

This cloud complex is dominated by the other two main submm peaks in the BLAST region, IRAS 19120+1103 (GRSMC 45.453+0.060; [G45.45]; a MYSO candidate in the RMS Survey; e.g., \citealp{mottram2010}) and IRAS 19117+1107 (GRSMC 45.478+0.131; [G45.48]). Both sources show significant extended submm emission (Figure \ref{fig:irac2}), and required additional Gaussians to model the emission local to the main submm peaks.  At \IRAS\ resolution the extended emission of both sources was not resolved, and so our BLAST flux densities were obtained with aperture photometry combining A28 with A29 (AA29; Figure \ref{fig:BB70irac}, comprising G45.47+0.05 and G45.45) and A30 with A30a (AA30).

\begin{figure}[ht]
\includegraphics[scale=0.50,angle=270]{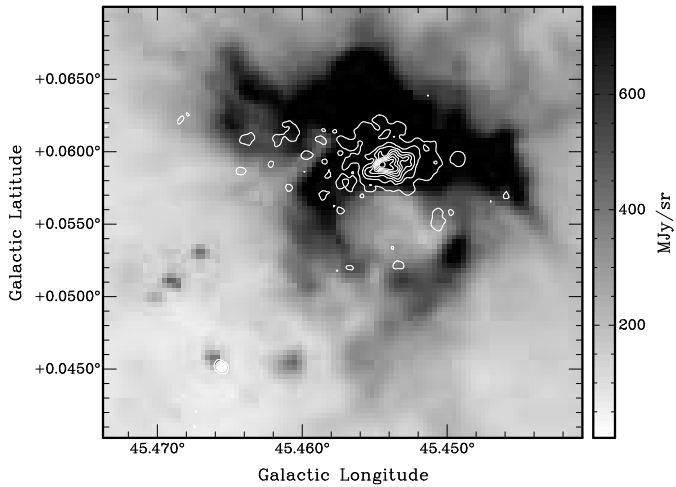}
\caption{Greyscale \irac\ 8\,\micron\ image of most central regions of A28 and A29 in Fig.\ref{fig:BB70irac} with CORNISH contours. Contours are from 10\% to 90\% of map peak value of 0.07\,Jy\,beam$^{-1}$ in 10\% steps.}
\label{fig:BB70iracbis}
\end{figure}

The presence of UCH{\sc ii}Rs was detected by WC89 and \citet{testi1999} at 3.6\,cm. \citet{chan1996} classified AA29 as a MYSO. The activity and complexity of this region is particularly prominent in the \irac\ images, as can be observed in Figures \ref{fig:irac2}-\ref{fig:BB70iracbis}.

\subsubsection{G45.47+0.05}
A29 (G45.47+0.05) is the most prominent of all submm peaks in this complex, and yet lacks significant emission in the MIR (e.g., Figure \ref{fig:BB70irac}; Figure \ref{fig:BB70iracbis}). The submm peak is located in the neighborhood of 14 GR sources (five of which are associated with bright IR sources; Figure \ref{fig:BB70irac}). All these sources are located within $\sim30$\arcsec\ of the BLAST peak. The appearance in the MIR may suggest that the submm emission is tracing a particularly dense part of the molecular clump. This material could be confining the expansion of the photodissociation region (PDR) as seen in \irac, which also appears to be in a very young and embedded stage of star formation. Pillar-like structures seen in extinction and extending from A29 into G45.45 (Figure \ref{fig:BB70irac}), and the main five IR detections embedded within the BLAST source (as suggested by their flux increasing with increasing wavelength) support this scenario.
 
\begin{figure}[ht]
\includegraphics[scale=0.45,angle=270]{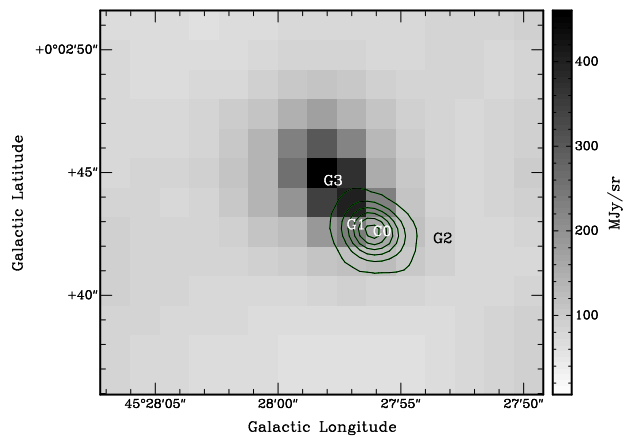}
\caption{Greyscale \irac\ 8\,\micron\ image of region around the main UCH{\sc ii}R (A29$-$C0; C0 in the image) within A29 (Figure \ref{fig:BB70irac}) with CORNISH contours superimposed. Contours are from 10\% to 60\% of map peak value of 0.07\,Jy\,beam$^{-1}$ in 10\% steps. Figure shows closest IR counterparts retrieved from the GLIMPSE I complete catalog (GC), labeled as G1, G2 and G3.}
\label{fig:irac2a}
\end{figure}

The brightest CORNISH source in the neighborhood of the submm peak, A16$-$C0 (Figure \ref{fig:irac2a}), is a known UCH{\sc ii}R (WC89, \citealp{testi1999} center agreement $<1$\arcsec). The flux from our radio analysis is equivalent to an O9.5 radio ZAMS star at 7\,kpc, which agrees with the spectral type we derived using the 6\,cm flux from \citet{urquhart2009}. This source is also one of the most complex of the clump, and contains H$_2$O (e.g., \citealp{forster1989}), OH (e.g., \citealp{forster1989}; \citealp{argon2000}) and methanol (e.g., \citealp{menten1991}) masers. The most prominent OH sources are located closer ($<0.5$\arcsec) to the radio peak than the methanol and H$_2$O masers ($\sim1$\arcsec), and they all lie preferentially towards the northern (upper left direction; Figure \ref{fig:irac2a}) rim of the radio emission and towards the extended MIR region. 

In the \irac\ maps, the GR source (G045.4661+00.0457; G3 in Figure \ref{fig:irac2a}) is at the peak of the IR emission, at the opposite corner of the MIR elongation to the CORNISH source (at $\sim3$\arcsec). Other GR sources are also present nearby the radio peak. Such astrometry differences between the radio and IR detections have been previously observed \citep{buizer2005}, which confirms this to be a true physical separation. The MIR elongation may be tracing the outflow morphology detected by \citet{hunter1997} and the NH$_3$ emission (e.g., \citealp{hofner1999}). Most maser sources are indeed located between the MIR peak (G3) and the radio peak, along the elongation, with a few OH masers located between the radio source and G2 (Figure \ref{fig:irac2a}). At the northern extreme edge of the elongation and closer to the main MIR peak there are more detections of methanol maser signatures \citep{kurtz2004}, as well as SiO and NH$_3$ emission peaks ($\sim$10-15\arcsec\ from the radio peak; \citealp{hofner1999}). The presence of shocked gas along the northern MIR emission, the outflow, and the location of the UCH{\sc ii}R opposite the main MIR peak may suggest the scenario where the outflow/jet originating at the UCH{\sc ii}R interacts with its local environment, producing the observed structure in the MIR. An alternative interpretation claims this elongation may arise from accretion \citep{buizer2005} or collapse (e.g., \citealp{cesaroni1992}; \citealp{hofner1999}) onto the UCH{\sc ii}R. This could explain the emission and redshifted absorption observed in NH$_3$. More recent studies have not confirmed the presence of infall signatures (e.g., \citealp{wilner1996}; \citealp{klaassen2007}). Further analysis of the innermost regions is needed to solve this controversy.

The maser emission (especially methanol and H$_2$O) suggests that this source is one of the youngest in this complex. It has also been classified as an Extended Green Object (EGO) and a likely MYSO outflow candidate \citep{cyganowski2008}. This supports the HCO analysis of \citet{wilner1996}, who suggested that this source is in the early stages of forming an OB star cluster.

\subsubsection{G45.45}
A28 is close to the most active region in the cloud as observed in the MIR (G45.45+0.06 [G45.45]), although from its position and extended morphology the submm emission may also be tracing the compressed and extended molecular gas. G45.45 (also a MYSO candidate in the RMS Survey; e.g., \citealp{mottram2010}) has been observed to lie in the border of a larger and fainter \ion{H}{2} region $\sim3$\arcmin\ in size (G45L; \citealp{paron2009}). Although this structure is also traced by the BLAST images, it has not been included in our photometry/radio analysis.

The \irac\ images show a complex morphology, with the strongest emission in the MIR and 21\,cm coincident with the main nebula-like radio structure observed in the CORNISH data (e.g., Figure \ref{fig:BB70irac}; Figure \ref{fig:BB70iracbis}). In Figure \ref{fig:BB70iracbis} the MIR emission appears to follow a horseshoe shape around a cavity-like structure ($\sim1$\,pc in size at 7\,kpc), within which no stars are detected in the \irac\ images. Although this structure can be observed at 1.3\,cm in the analysis of \citet{mooney1995} at $\sim12$\arcsec\ resolution, the details and morphology of the dust in this region remain unresolved in our BLAST maps. 

\begin{figure}[ht]
\includegraphics[scale=0.46,angle=270]{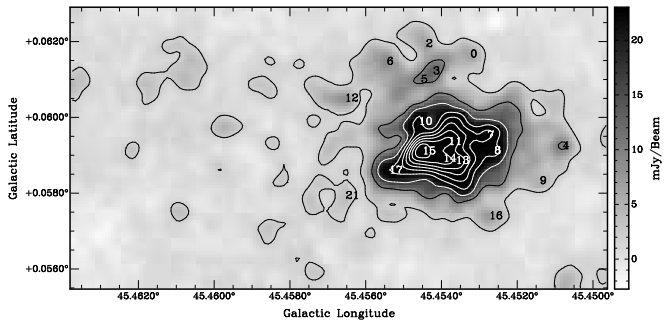}
\caption{Detailed greyscale CORNISH image of central UCH{\sc ii}R near A28 (part of IRAS 19120+1103; Fig.\ref{fig:BB70irac}; Fig.\ref{fig:BB70iracbis}), with numbered emission peaks (Table \ref{table:rsrcs}). Contours like Fig.\ref{fig:BB70iracbis}.}
\label{fig:cornish2b}
\end{figure}

\begin{figure}[t]
\includegraphics[scale=0.45,angle=270]{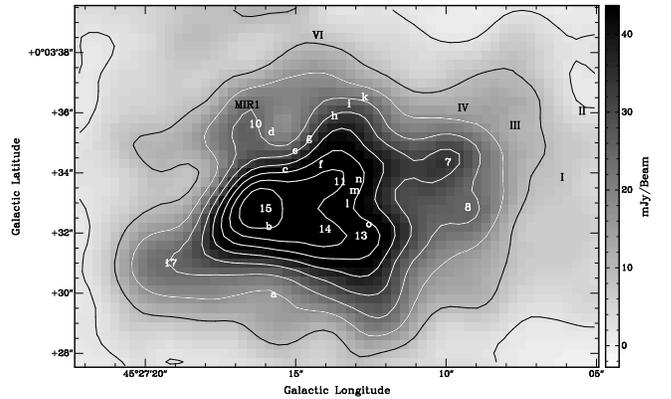}
\caption{More detailed greyscale CORNISH image of innermost regions of the central UCH{\sc ii}R in Fig.\ref{fig:cornish2b} with numbered emission peaks (Table \ref{table:rsrcs}). Letters are NIR detections from \citet{feldt1998} falling within VLA map (Table \ref{table:ir}). Roman numbers are other NIR detections \citep{feldt1998}. Contours like Fig.\ref{fig:BB70iracbis}.}
\label{fig:cornish2b2}
\end{figure}

In the radio, the CORNISH data reveal a similarly complex substructure for G45.45 (Figure \ref{fig:cornish2b}). There are 18 radio peaks detected within and surrounding the radio `nebula', coincident with the brightest area in the GLIMPSE images. Within this extended radio emission, called `The Orion nebula's younger brother' by \citet{feldt1998}, numerous emission peaks are noticeable. While we cannot easily distinguish or fit these peaks, we detect at least eight within an estimated size of $0.35\times0.2$\,pc at 7\,kpc (Figure \ref{fig:cornish2b2}). This supports previous studies that classify this source as a cluster of \ion{H}{2} regions (e.g., \citealp{garay1993}; \citealp{testi1999}). Aperture photometry on the whole structure (including extended emission) yields a radio flux density comparable to that of an $\sim$O5.5 ZAMS spectral type ($\sim$O5 class V) star, which is in agreement with previous estimates (e.g., \citealp{garay1993}). Using the 6\,cm flux from \citet{urquhart2009} for a source major axis of $\sim8$\arcsec, we obtain an estimate of $\sim$O6.5. This is comparable to the results we obtain using the flux estimates from Testi et al. (1999; Table \ref{table:radio1}).

Regarding the stellar content, \citet{feldt1998} detected up to 14 NIR sources (from \textit{a}-\textit{o}) possibly associated with this complex. They also detected additional mid-infrared (MIR) sources outside the main radio nebula (Figure \ref{fig:cornish2b2}). The roman numbers indicate other point sources in their NIR images outside their VLA map. We confirm their observations that the majority of these objects lie on the sharp and distorted upper edge of the radio emission (Figure \ref{fig:cornish2b2}). This may support the scenario proposed by these authors where a central cluster of OB stars, deeply embedded within the radio complex, produces a front that has triggered further star formation. In their analysis, they claim that IR sources \textit{l}, \textit{m}, \textit{n}, \textit{o} (Figure \ref{fig:cornish2b2}) may be the source of the UCH{\sc ii}R. Although these objects do lie close to two bright radio peaks within the cluster (A28$-$C13, A28$-$C11; Table \ref{table:ir}), their source \textit{b} is the closest to the brightest peak in the radio (A28$-$C15 in Figure \ref{fig:cornish2b2}). This CORNISH source lies at less than $\sim1$\arcsec\ from both the WC89 coordinates and source \textit{b}.

At a resolution of 0\arcsec.4, WC89 defined this source as a cometary UCH{\sc ii}R. Indeed, two parallel extensions appear to emerge from the central source A28$-$C15. These structures form a tail-like morphology extending towards the right in Figure \ref{fig:cornish2b2}, within which A28$-$C13, A28$-$C14, and A28$-$C11 are detected. In total, we find that eight of these IR sources have CORNISH counterparts within $\sim1$\arcsec. Sometimes one radio peak is found between several in the NIR (Table \ref{table:ir}; Figure \ref{fig:cornish2b2}). 

\citet{blum2008} classified the main radio structure as a group of MYSOs with late and early B stars surrounding the central source, probably similar to the central cluster powering G45.12 within A16. The offset between the NIR and radio peaks could be explained if the radio detections are dense clumps ionized by the surrounding massive stars \citep{feldt1998}. Alternatively, we do not discard the possibility that the NIR stars may have been produced by the central and embedded star/cluster, whose expanding \ion{H}{2} region could have triggered the formation of these sources by compression of the dense material traced in the submm. This last scenario is supported by the overall morphology in the CORNISH images; the observed emission supports A28$-$C15, the brightest radio source, as being responsible for the origin of \textit{l}, \textit{m}, \textit{n}, \textit{o}, rather than the latter sources being the actual `ground zero' for the triggering mechanism (as suggested by \citealp{feldt1998}). Further investigation at high resolution is required to corroborate the triggering hypothesis and to fully distinguish between the different cases.

\begin{table}[ht!]
\caption{4.8\,GHz radio (CORNISH) sources with IR counterparts \citep{feldt1998} within 1\arcsec.}
\label{table:ir}
\centering
\begin{tabular}{l l}
\hline
\hline
IR&Radio\\
\hline
b&A28$-$C15\\
d&A28$-$C10\\
f&A28$-$C11\\
l&A28$-$C11\\
m&A28$-$C11\\
n&A28$-$C11\\
o&A28$-$C13\\
MIR1&A28$-$C10\\
\hline
\end{tabular}
\end{table}

The candidate counterpart to MIR1 (A28$-$C10) is clearly visible in the CORNISH data, albeit relatively weak compared to the most central sources. This weak emission is in contrast with the strong IR luminosity of this source. In combination with the nearby OH maser \citep{argon2000}, this may again suggest that at least some structures are actual stars still embedded within their natal cocoons (e.g., \citealp{kraemer2003}).

The remaining possible radio detections within $\sim1$\arcmin\ and not lying within the main complex have been included in Table \ref{table:radio1}. Due to the complexity of the radio structures around this complex, we do not discard the presence of additional sources in this region (e.g., Figure \ref{fig:cornish2b}).

\subsubsection{G45.48}
AA30 (G45.48; north of AA29 in Figure \ref{fig:irac2}) also shows complex structure in the submm and IR. Our two submm peaks lie at opposite sides of the main emission visible in the \irac\ images. As in previous cases, they may be tracing relatively cold, dense gas that is confining the expansion of the \ion{H}{2} region.

 \begin{figure}[t]
\includegraphics[scale=0.57]{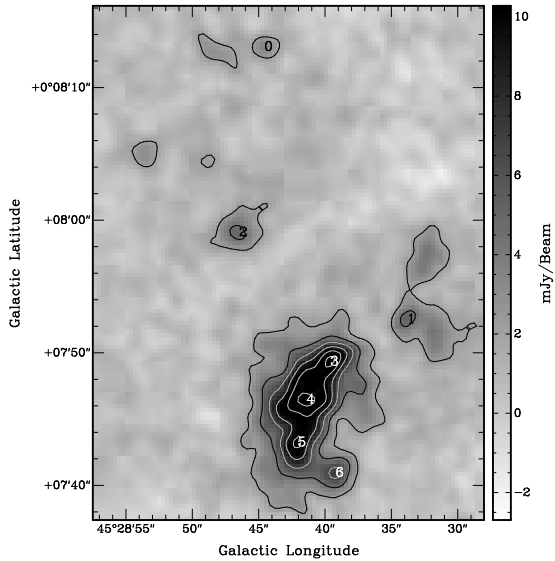}
\caption{Greyscale CORNISH image of elongated (and brightest) radio emission in A30$-$A30a (IRAS 19117+1107; Complex 2 in Table \ref{table:radio1}); with numbered emission peaks (Table \ref{table:rsrcs}). Contours are from 10\% to 25\% of map peak value of 0.07\,Jy\,beam$^{-1}$ in 3\% steps.}
\label{fig:cornish2c}
\end{figure}

\begin{figure}[t]
\includegraphics[scale=0.50,angle=270]{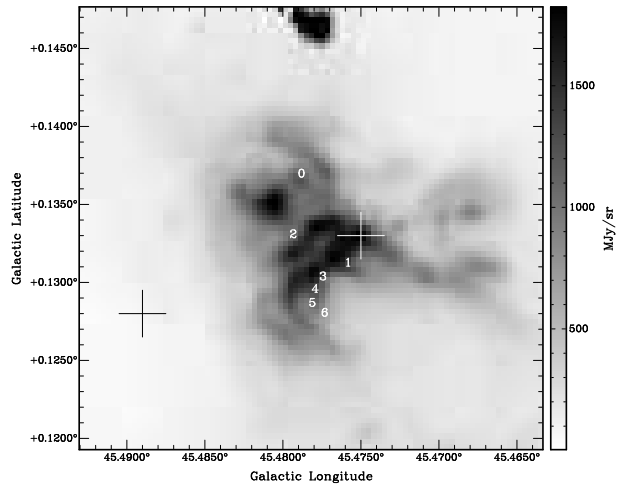}
\caption{Greyscale \irac\ 8\,\micron\ image of the \ion{H}{2} region within IRAS 19117+1107. Numbers as in Fig.\ref{fig:cornish2c}. Crosses mark the position of submm peaks A30 and A30a.}
\label{fig:irac2c}
\end{figure}

WC89 classified this UCH{\sc ii}R as irregular, but suggested that the elongated form could be due to an imaging artifact. \citet{testi1999} identified no compact sources, only extended emission. The elongation of this source, as well as several embedded peaks, are evident at 4.8\,GHz (Figure \ref{fig:cornish2c}), and the structure is identified in the MIR images as the equatorial ridge of the bipolar-like MIR structure (Figure \ref{fig:irac2c}).

Just like in G45.45, the MIR and main radio emission appear to wrap around a cavity-shaped structure, within which no \irac\ sources are visible. The BLAST emission includes this `cavity', but any such structure would be unresolved in our images. The adjacent radio/IR ridge could be compatible with illumination from the south, although the main source powering the region could in fact be located more towards the north, closer to source A30$-$C0 (Figure \ref{fig:irac2c}), the northern radio source detected by \citet{garay1993}, and the diffuse emission in the IR (KJK1) of \citet{kraemer2003}. Only one GC was found within $\sim5$\arcsec\ of A30$-$C0 (G045.4790+00.1365). KJK1 does appear to be coincident with an \irac\ emission peak, even though there is no counterpart in the GLIMPSE I catalogs. The irregular radio structures, and the lack of a significant radio counterpart for KJK1, could imply the presence of an embedded star/cluster around this position, which also coincides with an opacity peak and is estimated to be much colder than the southern counterpart \citep{kraemer2003}.

The multi-peaked radio morphology of the elongated radio structure (coincident with the southernmost source detected by \citealp{garay1993} and the IR source KJK2 from \citealp{kraemer2003}) clearly indicates the presence of embedded sources. Our Gaussian routine detects at least four main peaks in a filamentary structure about $\sim0.2$\,pc in length (Figure \ref{fig:cornish2c}). Aperture photometry on the entire filament yields a total radio flux equivalent to an O7.5 radio ZAMS (O8 class V) star. This is significantly later than the IR estimate (Table \ref{table:radio1}), perhaps an indication of self-absorption and a young embedded stage of evolution, despite our earlier spectral type estimate obtained using the 3.6\,cm flux from \citet{testi1999}. This structure is also coincident with methanol masers \citep{menten1991}, recognized as a signature of highly embedded stages of massive star formation. In addition, we observe (weaker) non-Gaussian, filamentary structures to the right of the main structure in Figure \ref{fig:cornish2c}, also likely associated with methanol emission. A prominent cluster of methanol, OH and water masers has been detected about $\sim15-20$\arcsec\ southwest from A30$-$C0 and A30$-$C1. The presence of a GC/GR counterpart (G045.4725+00.1335) for this source and the lack of an equivalent CORNISH detection again argues in favor of the young stage of the stellar population in this region.

\subsubsection{BLAST A35 \& A36}
\begin{figure}[ht]
\includegraphics[scale=0.50,angle=270]{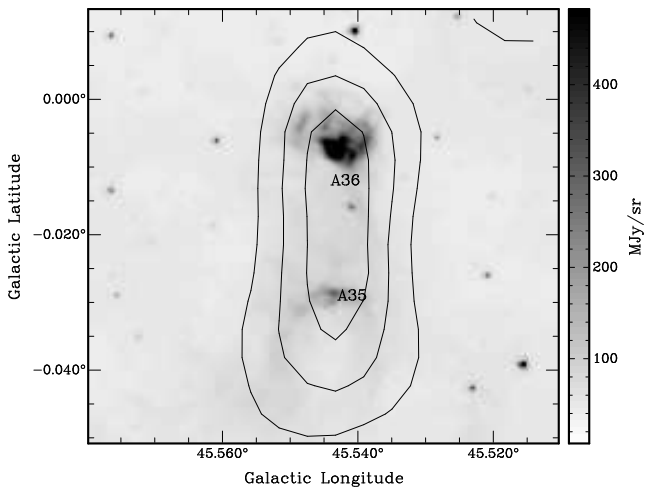}
\caption{Greyscale \irac\ 8\,\micron\ image showing cometary shape of A35+A36 with BLAST 500\,\micron\ contours superimposed. Contours are from 8\% to 12\% of map peak value of 2153\,MJy\,sr$^{-1}$ in 2\% steps.}
\label{fig:irac22}
\end{figure}

Away from the main complex, A35 and A36 (Figure \ref{fig:irac22}) also show some signatures of stellar activity, as suggested by their morphology. 

Although our BLAST peaks generally overlap well with the emission observed in \irac, \IRAS\ 100\,\micron\ shows poor correlation. Only at 60\,\micron\ do we observe a reasonable match with A36, the only one of these two sources with a PSC \IRAS\ match within $\sim30$\arcsec. This IR source, IRAS 19124+1106, was estimated to be at a distance of 6.5\,kpc, and has been suggested to be one of the impact sites of the jets emanating from the microquasar and superluminal source GRS 1915+105 (\citealp{kaiser2004}). The other impact site was identified as our BLAST source A19, a candidate MYSO from the Red MSX Source (RMS) Survey \citep{mottram2010}. GRS 1915+105 is believed to be a binary system consisting of an early-type K giant, accreting via a Roche lobe overflow onto a black hole (e.g., \citealp{greiner2001}). The MIR structure of A36, if associated with this \IRAS\ source, does not resemble what one would expect to see if caused by a jet originating at the position of GRS 1915+105. Furthermore, the new distance estimate for GRS 1915+105 of 11\,kpc \citep{zdziarski2005} is now much larger than the kinematic distance to this IR/submm emission. 

Our analysis of the CORNISH images reveals that, with the exception of some weak structures near the emission peak of A36 (likely tracing the ionized region as observed in the VGPS 21\,cm maps), there are no major radio sources near these objects. This may suggest very early or very late/isolated low mass star forming activity.

\subsubsection{Submm Analysis}
Estimated BLAST parameters for the main objects in this complex have been included in Table \ref{table:seds}. The temperatures are warm, characteristic of active star forming clumps with ongoing stellar activity. The total mass of AA29 is $\sim4800$\,M$_\odot$ for a temperature of $\sim35$\,K, which contrasts with the mass estimate of 26000\,M$_\odot$ obtained by \citet{mooney1995} for a distance of 8.1\,kpc and a much lower fixed dust temperature of 20\,K. Scaling to this distance we obtain a total mass of $\sim6300$\,M$_\odot$, and $\sim34000$\,M$_\odot$ for a fixed temperature of 20\,K and free $\beta$ (best fit $\beta=2.2$). This shows the importance of actually determining the temperature using the BLAST observations.

Our estimated IR+submm single-star spectral type for AA29 of $\sim$O5.5 is in excellent agreement with the radio spectral type of the radio nebula within G45.45 (Complex 1 in Table \ref{table:radio1}), the most luminous system within the BLAST emission formed by A28 and A29 (AA29).

The IR+submm spectral type of A30+A30a is in agreement with the IR spectral type estimate of Complex 2 (the elongated radio emission; Table \ref{table:radio1}), but the CORNISH radio spectral type of this substructure is considerably later. 

Although the radio flux estimates from \citet{testi1999} at 3.6\,cm and the uncertainties in the SED bolometric luminosities could be compatible with a relatively optically thin state for this source, the CORNISH measurements suggest a more embedded (and younger) stage for AA30 than for AA29. Just like the clumps within GRSMC G045.14+00.14, the BLAST sources in GRSMC G045.49+00.04 appear to be powered by the central OB young stellar clusters, which could also be responsible for the additional star formation detected within their parent clumps.

\subsection{GRSMC G045.74$-$00.26 \& GRSMC G045.89$-$00.36}
This complex is dominated by three main (sub)mm sources (Figure \ref{fig:irac3}) within an extended and largely filamentary molecular region.
\subsubsection{BLAST A50}
In the submm, A50 (IRAS 19145+1116) is largely structureless and symmetric, possibly with some weak emission surrounding the central peak. In the  MIR, this \ion{H}{2} region reveals a clumpy structure and numerous MIR peaks clustered within the main submm emission. There are seven GC sources found at less than $\sim20$\arcsec\ from our BLAST peak. Among the closest of these we find a string of three sources, G045.9352$-$00.4017 (the central one), G045.9362$-$00.4017 (the brightest), and G045.9344$-$00.4016, which constitute the most prominent structure in the GLIMPSE images.

Overall, the extended structure in the MIR and at 21\,cm is rather circular, although emission arcs are visible within the region in the MIR. We detect only one weak 4.8\,GHz source, found at less than $\sim3$\arcsec\ from the GLIMPSE source G045.9344-00.4016, with a flux corresponding to a B0.5$-$B0 ZAMS star. The IRAS clump was reported to satisfy the color criteria of WC89 for UCH{\sc ii}Rs (e.g., \citealp{bronfman1996}), and was reported by \citet{codella1995} to be possibly associated with water masers. This would argue against the more evolved state of the system indicated by its `disturbed' morphology, although from the coordinates provided by these authors we suspect that this maser emission could also be associated with a later generation of star formation within an older clump. The lack of a CORNISH or IR counterpart for this maser may also be indicating that the structure is due to a low mass star.

\begin{table*}[ht!]
\caption{Masses and temperatures of the main BLAST sources at $\sim7$\,kpc.}
\label{table:seds}
\centering
\begin{tabular}{l l l l l}
\hline
\hline
Source&$T$&$M_{\mathrm{c}}$&$L_{\mathrm{bol}}$&Single-star Spectral Type\\
&(K)&($10^2$\,M$_{\odot}$)&($10^3$\,L$_{\odot}$)&\\
\hline
A12&$42.5\pm0.5$&$14.2\pm1.0$&$282.8\pm0.6$&O6\tablenotemark{a}/O5\tablenotemark{b}\\
A16&$39.9\pm4.6$&$35.4\pm8.0$&$340.9\pm131.7$&O6$-$O5.5/O6$-$O4\\  
A22\tablenotemark{c}&$16.2\pm4.6$&$65.7\pm47.6$&$7.5\pm6.6$&$<$B0.5/$-$\\ 
A28$-$A29&$36.7\pm4.8$&$47.8\pm10.3$&$262.3\pm121.6$&Combined due to \IRAS\ unresolved (and saturated) emission\\
&&&&O6.5$-$O5.5/O7$-$O4.5\\
A30$-$A30a&$27.4\pm5.8$&$33.4\pm9.1$&$89.1\pm67.2$&Combined due to \IRAS\ unresolved (and saturated) emission \\
&&&&B0$-$O6.5/$<$O6.5\\
A35&$33.9\pm10.9$&$4.3\pm1.5$&$62.0\pm144.9$&$\sim$O8/$\sim$O9\\
A36&$41.6\pm1.4$&$1.7\pm0.3$&$30.1\pm0.4$&B0/$-$\\
A46&$32.5\pm1.6$&$4.5\pm0.7$&$19.1\pm2.7$&B0.5$-$B0/$-$\\
A47&$37.9\pm2.5$&$7.9\pm1.3$&$76.0\pm14.7$&O8$-$O7.5/O9$-$O8\\
A50&$35.0\pm2.0$&$6.9\pm1.2$&$43.6\pm6.6$&O9.5$-$O8.5/$<$O9.5\\
\hline
\multicolumn{5}{l}{{$^a$ Spectral type from L$_{\mathrm{bol}}$ for a ZAMS star \citep{panagia1973}.}}\\
\multicolumn{5}{l}{{$^b$ Spectral type from L$_{\mathrm{bol}}$ for a luminosity class V star \citep{martins2005}.}}\\
\multicolumn{5}{l}{{$^c$ No PSC counterpart within 30\arcsec. No \bolocam\ data available. Possibly associated (visually)}}\\
\multicolumn{5}{l}{{with an \IRAS\ source, but this source covers nearby BLAST sources as well and is therefore ignored in the}}\\
\multicolumn{5}{l}{{fits. The rising SED in the submm and the lack of data at shorter wavelengths result in large error ranges.}}\\
\end{tabular}
\end{table*}

\subsubsection{BLAST A46}
While relatively structureless in the submm, the GLIMPSE images of A46 (IRAS 19141+1110) show a bow-shock cometary like structure, with the front oriented towards the south (lower right of Figure \ref{fig:irac3}). Some possibly associated sources towards the south and southeast of the main structure are also detected. Despite the presence of numerous GC/GR sources, the bow-shock itself may be produced by a highly embedded source, closest to G045.8062-00.3518 (Figure \ref{fig:irac3}).

Even though A46 also has \IRAS, \textit{MSX}, and \irac\ counterparts, there appears to be negligible 21\,cm emission and no NVSS, 4.8\,GHz emission, H$_2$O or OH masers, or any other tracers that may be indicative of ongoing star formation. An exception is the methanol maser emission $\sim25$\arcsec\ eastwards (lower left) from the BLAST peak detected by \citet{pandian2007a}. This is the 5$_1-6_0$A$^+$ line of methanol at 6.7\,GHz, the strongest of Class II methanol masers, seen exclusively near massive stars. The authors suggest a very early stage of evolution, in which methanol activity has just turned on, before the onset of additional signs of star formation. Class II maser sources are radiatively pumped but are believed to be quenched by the formation of an UCH{\sc ii}R, which is consistent with the absence of 4.8\,GHz CORNISH sources and the `young' age of the system (e.g., \citealp{minier2005}; \citealp{ellingsen2007}). This scenario would also be in agreement with \citet{cyganowski2008}, who classified this source as a `possible' MYSO. The observed \irac\ structures and methanol emission are all within the prominent BLAST submm emission, which also suggests the presence of local high density molecular material. 

\subsubsection{BLAST A47}
\begin{figure}[t]
\includegraphics[scale=0.49,angle=270]{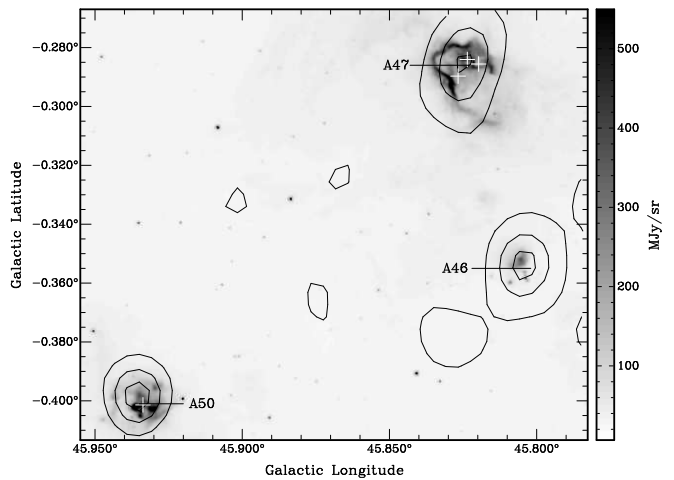}
\caption{Greyscale \irac\ 8\,\micron\ image of IRAS 19141+1110 (A46), IRAS 19139+1113 (A47), and IRAS 19145+1116 (A50), with BLAST 500\,\micron\ contours superimposed. Contours are from 10\% to 20\% of map peak value of 2153\,MJy\,sr$^{-1}$ in 5\% steps. White crosses are CORNISH detections.}
\label{fig:irac3}
\end{figure}

A47 (IRAS 19139+1113) is a prominent MYSO (e.g., \citealp{chan1996}) showing clear signatures of active star formation. A prominent carved structure ($\sim2$\,pc in size at 7\,kpc) is observed towards the north (Figure \ref{fig:irac3}). This \ion{H}{2} region is the most extended feature of this cloud, and is detectable in the IR, submm, mm, and 21\,cm images. The submm and radio peaks appear to trace a region of diffuse emission within this structure, close to three bright \irac\ sources (G045.8263$-$00.2820, G045.8212$-$00.2846, and G045.8291$-$00.2858). The location of these IR objects, close to the center of curvature of some of the ionization fronts forming the edge of the ionized region in the molecular cloud, suggests that the combined contribution from these objects results in the sinusoidal boundary of the overall region. 

We find three likely sources at 4.8\,GHz within $\sim1$\arcmin\ of our BLAST source, all with GC counterparts within $\sim5$\arcsec, and indicative of ongoing massive star formation. The extended morphology and the advanced state of stellar activity within the main \ion{H}{2} would support the `older' age suggested by the lack of maser emission (e.g., \citealp{walt1995}; \citealp{szymczak2000}).

\subsubsection{Radio and Submm Analysis}
From their measured fluxes, we estimate ZAMS spectral types for the radio peaks in these clumps of later than $\sim$B0 stars. Our SED fits show that, overall, none of the host clumps have an equivalent single-star ZAMS spectral type earlier than an $\sim$O8 star. In the case of A47, it is possible that additional sources are contributing to the total luminosity, in addition to those traced in the radio. 

The detection of 4.8\,GHz emission confirms previous studies that classified A47 and A50 as UCH{\sc ii}Rs from the WC89 color criteria \citep{bronfman1996}. Although the radio peaks are not coincident with the brightest \irac\ sources, they clearly indicate that massive stars are indeed being formed in this cloud, albeit at a lower level than that observed in the two main clouds described above. 

\section{Aquila in Perspective: The Star Forming Activity In the BLAST 6 deg$^2$ Map} \label{sec:discussion}
\subsection{The Submm Population and Structural Properties of the Aquila Field}
This region of Aquila consists principally of three main active complexes at $\sim$7\,kpc (G045.49+00.04, G045.14+00.14, and the neighboring region of G045.74$-$00.26 and G045.89$-$00.36), as well as a quiescent cloud (GRSMC 45.60+0.30) at 1.8\,kpc (e.g., \citealp{simon2001}).  There also exists (at various distances) additional dispersed (clump) star forming activity throughout the field, two cataloged (and two candidate) supernova remnants \citep{green2009}, and a microquasar (GRS 1915+105). 

In the present work we have used our BLAST submm data to identify and characterize the global parameters of the clumps at $\sim7\pm0.5$\,kpc, in which the molecular emission (and star formation activity) is the strongest. These clumps have been further characterized in the MIR and the radio with \irac\ GLIMPSE and 21\,cm continuum maps. The stellar content as well as the molecular structure have been investigated with CORNISH interferometry data and GRS $^{13}$CO/CS emission.

At $\sim60\pm10$\,km s$^{-1}$ the $^{13}$CO maps reveal a dispersed and filamentary molecular structure covering the size of the BLAST map. Two of the main complexes are located at the central `knots' of an overall heart-shaped like morphology (Figure \ref{fig:molecular}). From the catalog of \citet{rathborne2009} and \citet{roman2009} only five clouds are at a distance of $\sim7\pm0.7$\,kpc (six including G045.14+00.14), but these contain the brightest objects in the field. 

The submm emission follows the molecular emission well (Figure \ref{fig:molecular}). Clumps with typical star forming characteristics ($T\sim30-35$\,K for $\beta=1.5$) are associated mainly with the three main complexes. Outside of the central regions the star forming activity as traced in the submm is less organized, although still following the overall contours of the molecular emission. More `isolated' clumps with \IRAS\ counterparts (not all part of our final 7\,kpc sample) show typical star-forming clump temperatures of $\sim30$\,K and disturbed environments, with their stellar content generally resolved, and in an apparently less embedded stage of evolution. 

Relatively cold structures in Aquila (T$<20$\,K for $\beta=1.5$ as measured in this analysis) have counterparts in major IRDC surveys. \citet{simon2006} detected 20 IRDC candidates in absorption with \textit{MSX}, and \citet{peretto2009} found 79 candidates using \textit{Spitzer} data. We have checked for IRDC matches within 30\arcsec\ of our submm peaks in the entire field and found seven examples, all from the sample of \citet{peretto2009}. We note that several IRDCs, although not located within 30\arcsec\ of our submm peaks, are frequently clustered in their immediate neighborhood; these may have not be resolved in our data, or they may be confused by deconvolution structures. One prominent example of this is G045.49+00.04. Several IRDCs are observed in the local regions surrounding the brightest \IRAS\ source formed by A28 and A29, especially towards A29. The intricate structures observed in the \irac\ images, such as the pillar-like structures coincident with this BLAST source and the `cavity', will require submm data with higher resolution and sensitivity in order to be properly characterized. We find a similar situation for the submm complex labeled in Figure \ref{fig:molecular} as G046.34$-$00.21, where the majority of the submm sources are identified as dark features in the \irac\ maps. Many of these objects appear to be part of globular-like structures `blown' by the stellar activity located to the northeast. The characteristics of these and similar sources outside the distance range chosen for the present work (which focuses on the most prominent and active clumps of the field) will be investigated in detail with the release of new \textit{Herschel} data from Hi-Gal.

Although a more in depth study is required to confirm our results, the distribution and characteristics of the population within the main clumps suggest that triggering mechanisms (new star formation induced/initiated by external agents; e.g., \citealp{elmegreen1992}) are likely taking place in the region. Our multiwavelength (MIR-radio) analysis is compatible with induced activity from the innermost region to the farthest (parsec) extent of the parent clump.

\subsection{Evolution and Massive Star Formation in Aquila}

\begin{figure}[ht]
\includegraphics[scale=0.48]{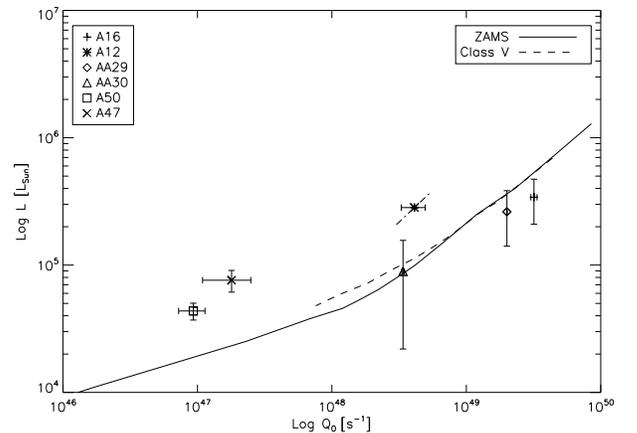}
\caption{Comparison of L$_{\mathrm{bol}}$ (Table \ref{table:seds}) and total Q$_0$ for BLAST sources with analyzed CORNISH sources (Tables \ref{table:radio} \& \ref{table:radio1}). Results are compared with theoretical curves for ZAMS \citep{panagia1973} and Class V \citep{martins2005} stars. Example of expected effect of increasing (decreasing) the distance by 1 kpc shown by diagonal line of source A12.}
\label{fig:check}
\end{figure}

Within G045.14+00.14, our analysis in the IR, submm, and radio, the molecular data, and the maser and stellar activity all support the scenario where A16 is in a more evolved stage of evolution than its neighbor A12. Figure \ref{fig:check} shows the measured bolometric luminosity for our BLAST sources relative to the combined total number of ionizing photons per unit time for all CORNISH sources possibly associated with each submm source. The apparent underionizing state of A12 suggests optically thick radio emission, and therefore a highly embedded stage of the central sources.

A28 likely contains the `older' stellar activity of the observed clumps, while A29 and A30 appear to have characteristics of a relatively `younger' population. The later spectral type obtained from the CORNISH data for the main complex in A30 (the filamentary structure) relative to its estimate from the IR would support the embedded stage of the main star forming activity in this region. Although our total luminosity measurement (and the radio estimates from \citealp{testi1999}) are consistent with the theoretical curve in Figure \ref{fig:check}, the large error range in luminosity for AA30 relative to the uncertainty in $Q_0$ could still support this conclusion.

Despite the apparent agreement of AA29 (A28+A29) with the theoretical models in Figure \ref{fig:check}, the more compact radio emission of A29 and the weak MIR emission (compared to the prominent submm peak) would argue that A29 is the youngest active part of the entire cloud. \citet{paron2009} suggested that G45.45 itself could have been triggered by three possible O-type stars ionizing the extended \ion{H}{2} region G45L. Although not analyzed in the present work, this is also a possibility for the original site (`first generation') of the activity within the BLAST clump. 

A35 and A36 also have SED parameters consistent with on-going stellar activity, but lack obvious 4.8\,GHz detections. Neither of these two sources show maser emission in the catalogs used in this work, and both have weaker $^{13}$CO and CS antenna temperature. These sources may thus be in a relatively young and still embedded stage, or the star forming activity may be low and/or without significant massive star formation. 

Of the three main submm peaks in the complex formed by G045.74$-$00.26 and G045.89$-$00.36, A50 has the strongest molecular emission. It is also possibly associated water masers, and has a relatively symmetric IR, submm and radio configuration. This source appears to have at least one possibly associated 4.8\,GHz source, which would agree with previous studies that classified this source as an UCH{\sc ii}R from the color criteria of WC89.

A46 shows the weakest molecular and radio emission, and no CORNISH counterparts or prominent signatures of star forming activity. However, the presence of methanol emission and the possible association with an EGO suggest that this clump is in a very early and deeply embedded stage of massive star formation.

A47 shows not only more extended emission at all wavelengths, but it also appears to contain a few associated CORNISH sources within the main \ion{H}{2} region. The lack of any obvious maser emission, its morphology, the high dust temperatures, and weak molecular emission indicate that this clump has the most evolved stellar activity of the entire complex. The weak $^{13}$CO and CS antenna temperatures for this source could also be explained by the more evolved stage of the system.

From the results in Figure \ref{fig:check}, both A47 and A50 appear to be underionizing for their luminosity. This could be an indication that the total luminosity has significant contribution from several `cooler' stars. However, the presence of water masers and CORNISH detections may still suggest that unaccounted very optically thick radio emission, from a new embedded generation of star formation in these clumps, could also be a possibility.

\section{Conclusion} \label{sec:end}
By means of the submillimeter maps provided by BLAST at 250, 350, and 500\,\micron, together with \irac\ imaging, 21\,cm continuum emission, 4.8\,GHz CORNISH radio interferometry data, and GRS FCRAO $^{13}$CO/CS molecular datacubes, we have characterized the main clump population in the region of Aquila.

Our SED fitting of the most prominent star-forming sources at a distance of $\sim7$\,kpc has revealed clumps with temperatures ranging between $T\sim35$\,K and $40$\,K (for $\beta=1.5$). Their total bolometric luminosities have equivalent single-star ZAMS spectral types earlier than $\sim$B0 stars.

The 4.8\,GHz interferometry maps have been used to investigate the UCH{\sc ii}Rs within the submm clumps. The ZAMS spectral types estimated from the CORNISH data are overall in good agreement with previous results. The `later' types obtained for some sources relative to their IR estimates support our BLAST and maser/outflow analysis, which suggest that highly embedded young OB stellar clusters are powering the most active clumps in the region.
 
On-going massive star formation is most prominent in G045.49+00.04 and G045.14+00.14 (containing our BLAST sources A12, A16, and A28, A29, A30, respectively). We confirm the presence of an OB stellar cluster deep within IRAS 19111+1048 (A16), containing a bright $\sim$O6 star surrounded by numerous late O and early B stars. This is in contrast with the scarcer maser and 4.8\,GHz emission towards the complex formed by G045.74$-$00.26 and G045.89$-$00.36, in the field containing A46, A47, and A50. Despite the `disturbed' environments observed in the MIR, the evolutionary stage of the three main clumps in this complex appear to range from quite evolved (with extended and bright rimmed \ion{H}{2} regions and a prominent IR stellar population), to the earliest stages of massive star formation detected so far in this field (A46, prior to the onset of significant maser emission and UCH{\sc ii}Rs). The parameters derived from these SEDs are typical of active clumps and are indeed suggestive of massive star formation, although at a lower level to that observed in the previous two clouds.

The enhanced datasets that will be provided by \textit{Herschel} (Hi-GAL) and SCUBA-2 at JCMT (JCMT Galactic Plane Survey; \citealp{moore2005}) will be crucial to fully characterize and complete the census of the clump population in this field, especially the fainter and the coldest structures. In addition, the higher resolution of these instruments will help to probe down to core scales and to investigate the dusty structures within the clumps, which appear to be channeling the stellar activity and shaping the ionized regions in the neighborhood of the embedded OB stellar clusters.

\acknowledgements The BLAST collaboration acknowledges the support of NASA through grants NAG5-12785, NAG5-13301, and NNGO-6GI11G, the Canadian Space Agency (CSA), the UK Particle Physics and Astronomy Research Council (PPARC), the Canada Foundation for Innovation (CFI), the Ontario Innovation Trust (OIT), and Canada's Natural Sciences and Engineering Research Council (NSERC). We would also like to thank the Columbia Scientific
Balloon Facility (CSBF) staff for their outstanding work. We also thank the referee for very useful suggestions and improvements to our paper, and Mubdi Rahman for useful discussions.

\bibliographystyle{apj}

\end{document}